\documentclass[fleqn,usenatbib]{mnras}

\usepackage{newtxtext,newtxmath}

\usepackage[T1]{fontenc}
\usepackage{ae,aecompl}
\usepackage{natbib}

\def\lya{Ly$\alpha$}

\def\fesc{\ifmmode f_{\rm esc} \else $f_{\rm esc}$\fi}

\def\msun{M$_\odot$}

\def\oiiil{[O~{\sc iii}]$\lambda 5007$}
\def\oiiill{[O~{\sc iii}]$\lambda 5007$}

\usepackage{graphicx}	
\usepackage{amsmath}	
\usepackage{amssymb}	

\title[LyC leaking galaxies with high O$_{32}$]{Low-redshift 
Lyman continuum leaking galaxies with high [O~{\sc iii}]/[O~{\sc ii}] ratios}

\author[Y. I. Izotov et al.]{
Y. I. Izotov$^{1,2,3}$, 
G. Worseck$^{4}$, 
D. Schaerer$^{5,6}$, 
N. G. Guseva$^{1,2,3}$, 
T. X. Thuan$^{7}$, 
\newauthor 
~K. J. Fricke$^{8,3}$, A. Verhamme$^{5}$ \& I. Orlitov\'a$^{9}$ 
\\
$^{1}$Main Astronomical Observatory, National Academy of Sciences of Ukraine,
27 Zabolotnoho str., Kyiv 03143, Ukraine,\\ 
E-mail: izotov@mao.kiev.ua, guseva@mao.kiev.ua\\
$^{2}$Bogolyubov Institute for Theoretical Physics,
National Academy of Sciences of Ukraine, 14-b Metrolohichna str., Kyiv,
03143, Ukraine\\
$^{3}$Max-Planck-Institut f\"ur Radioastronomie, Auf dem H\"ugel 69, D-53121 
Bonn, Germany\\
$^{4}$ Institut f\"ur Physik und Astronomie, Universit\"at Potsdam, Karl-Liebknecht-Str. 24/25, D-14476 Potsdam, Germany,\\
E-mail: gworseck@uni-potsdam.de\\
$^{5}$Observatoire de Gen\`eve, Universit\'e de Gen\`eve, 
51 Ch. des Maillettes, 1290, Versoix, Switzerland,\\
E-mail: daniel.schaerer@unige.ch, anne.verhamme@unige.ch\\
$^{6}$IRAP/CNRS, 14, Av. E. Belin, 31400 Toulouse, France\\
$^{7}$Astronomy Department, University of Virginia, P.O. Box 400325, 
Charlottesville, VA 22904-4325, USA,\\
E-mail: txt@virginia.edu\\
$^{8}$Institut f\"ur Astrophysik, G\"ottingen Universit\"at, 
Friedrich-Hund-Platz 1, D-37077 G\"ottingen, Germany,\\
E-mail: kfricke@gwdg.de\\
$^{9}$Astronomical Institute, Czech Academy of Sciences, Bo\v cn{\'\i} II 1401, 
141 00, Prague, Czech Republic,\\
E-mail: orlitova@asu.cas.cz}

\date{Accepted XXX. Received YYY; in original form ZZZ}

\pubyear{2017}

\begin{document}
\label{firstpage}
\pagerange{\pageref{firstpage}--\pageref{lastpage}}
\maketitle

\begin{abstract}
We present observations with the Cosmic 
Origins Spectrograph onboard the {\sl Hubble Space Telescope} of
five star-forming galaxies at redshifts $z$ in the range 0.2993 -- 0.4317
and with high emission-line flux ratios O$_{32}$ = 
[O~{\sc iii}]$\lambda$5007/[O~{\sc ii}]$\lambda$3727 $\sim$ 8 -- 27
aiming to detect the Lyman continuum (LyC) emission. We detect LyC emission in 
all galaxies with the escape fractions $f_{\rm esc}$(LyC) 
in a range of 2 -- 72 per cent. 
A narrow Ly$\alpha$ emission line with two peaks in four galaxies and with
three peaks in one object is seen in medium-resolution COS spectra
with a velocity separation between the peaks $V_{\rm sep}$ varying 
from $\sim$153~km~s$^{-1}$ to $\sim$ 345 km s$^{-1}$.
We find a general increase of the LyC escape fraction with increasing
O$_{32}$ and decreasing stellar mass $M_\star$, but with a large scatter of 
$f_{\rm esc}$(LyC). A tight anti-correlation is found 
between $f_{\rm esc}$(LyC) and $V_{\rm sep}$ making $V_{\rm sep}$ a good parameter
for the indirect determination of the LyC escape fraction. 
We argue that one possible source driving the escape of ionizing radiation is  
stellar winds and radiation from hot massive stars.
\end{abstract}




\section{Introduction}\label{intro}

Much efforts have been made in recent years to identify the sources of the
reionization of the Universe at redshifts $z$ = 5 -- 10. Two competitive
possible sources of ionization, active galactic nuclei (AGN) \citep*{Madau15} 
and low-mass star-forming galaxies (SFGs) \citep*{O09,WC09,M13,Y11,B15a},
were proposed. While there are pro and con arguments on the role of AGN
in the reionization of the Universe \citep*[e.g. ][]{Madau15,H18,M18,P18}, their
contribution remains uncertain. On the other hand, SFGs can be considered
as an important source of ionization only if the escape fraction of their
ionizing radiation on average is not less than 10--20 per cent 
\citep[e.g.][]{O09,R13,D15,Robertson15,K16}.

Currently, only few reliable Lyman continuum (LyC) leakers with 
high escape fraction
are known at high redshifts, {\em Ion2} at $z=3.218$ \citep{Va15,B16},
Q1549-C25 at $z=3.15$ \citep{Sh16}, A2218-Flanking at $z\approx 2.5$ 
\citep{B17}, {\em Ion3} at $z=3.999$ \citep{Va18}. 
The difficulty in the identification of high-$z$
LyC leakers is caused by their faintness, contamination by lower-redshift
interlopers and attenuation by partially neutral intergalactic medium (IGM) 
\citep[e.g., ][]{V10,V12,Inoue14,Gr15}. To overcome this 
difficulty it was proposed to study local compact low-mass SFGs as
proxies of high-$z$ galaxies \citep[e.g. ][]{I16,I16b}.
Their stellar masses, metallicities and star-formation rates are similar
to those of high-$z$ Lyman-alpha emitting (LAE) galaxies \citep{I15,I16c}.
A fair fraction of compact SFGs is characterised by high line ratios 
O$_{32}$ = [O~{\sc iii}]$\lambda$5007/[O~{\sc ii}]$\lambda$3727 $\ga$ 5 
\citep{S15} indicating that they may contain density-bounded H~{\sc ii} regions 
and be potential LyC leakers as suggested e.g.\ by \citet{JO13} and 
\citet{NO14}.

\citet{I16,I16b,I18}
selected six compact SFGs at redshifts $\sim$ 0.3 with high O$_{32}$ $\ga$ 5
for observations with the {\sl Hubble Space Telescope} 
({\sl HST})/Cosmic Origins Spectrograph (COS).
They showed that all six galaxies are LyC leakers with
$f_{\rm esc}$(LyC) ranging between 6 and 46 per cent. These values
are much higher than $f_{\rm esc}$(LyC) of $\sim$0 -- 4.5 per cent derived 
e.g. by 
\citet{B14} and \citet{C17}
in SFGs with lower O$_{32}$
or with lower equivalent widths of the H$\beta$ emission line \citep{He18}.

  \begin{table*}
  \caption{Coordinates, redshifts, distances and O$_{32}$ ratios of selected 
galaxies
\label{tab1}}
\begin{tabular}{lrrccrr} \hline
Name&R.A.(2000.0)&Dec.(2000.0)&$z$&$D_L$$^{\rm a}$&\multicolumn{1}{c}{$D_A$$^{\rm b}$}&O$_{32}$ \\ \hline
J0901$+$2119&09:01:45.61&$+$21:19:27.78&0.2993&1562& 925& 8.0\\
J1011$+$1947&10:11:38.28&$+$19:47:20.94&0.3322&1763& 994&27.1\\
J1243$+$4646&12:43:00.63&$+$46:46:50.40&0.4317&2401&1172&13.5\\
J1248$+$4259&12:48:10.48&$+$42:59:53.60&0.3629&1956&1053&11.8\\
J1256$+$4509&12:56:44.15&$+$45:09:17.01&0.3530&1893&1034&16.3\\
\hline
\end{tabular}

\hbox{$^{\rm a}$Luminosity distance in Mpc \citep[NED, ][]{W06}.}

\hbox{$^{\rm b}$Angular size distance in Mpc \citep[NED, ][]{W06}.}
  \end{table*}

  \begin{table*}
  \caption{Apparent AB magnitudes with errors in parentheses compiled
from the SDSS, {\sl GALEX} and {\sl WISE} databases
\label{tab2}}
\begin{tabular}{lccccccccccccc} \hline
Name&\multicolumn{5}{c}{SDSS}
&&\multicolumn{2}{c}{\sl GALEX}&&\multicolumn{4}{c}{{\sl WISE}} \\ 
    &\multicolumn{1}{c}{$u$}&\multicolumn{1}{c}{$g$}&\multicolumn{1}{c}{$r$}&\multicolumn{1}{c}{$i$}&\multicolumn{1}{c}{$z$}&&FUV&NUV&&\multicolumn{1}{c}{$W1$}&\multicolumn{1}{c}{$W2$}&\multicolumn{1}{c}{$W3$}&\multicolumn{1}{c}{$W4$}
\\
    &\multicolumn{1}{c}{(err)}&\multicolumn{1}{c}{(err)}&\multicolumn{1}{c}{(err)}&\multicolumn{1}{c}{(err)}&\multicolumn{1}{c}{(err)}&&(err)&(err)&&\multicolumn{1}{c}{(err)}&\multicolumn{1}{c}{(err)}&\multicolumn{1}{c}{(err)}&\multicolumn{1}{c}{(err)} \\
\hline
J0901$+$2119& 22.01& 21.54& 20.31& 21.94& 20.33&& 22.06& 22.17&&\multicolumn{1}{c}{ ... }&\multicolumn{1}{c}{ ... }&\multicolumn{1}{c}{ ... }&\multicolumn{1}{c}{ ... } \\
            &(0.16)&(0.05)&(0.03)&(0.15)&(0.12)&&(0.12)&(0.14)&&\multicolumn{1}{c}{(...)}&\multicolumn{1}{c}{(...)}&\multicolumn{1}{c}{(...)}&\multicolumn{1}{c}{(...)} \\
J1011$+$1947& 21.71& 21.26& 19.82& 21.64& 20.52&& 22.08& 21.31&&\multicolumn{1}{c}{ ... }&\multicolumn{1}{c}{ ... }&\multicolumn{1}{c}{ ... }&\multicolumn{1}{c}{ ... } \\
            &(0.13)&(0.05)&(0.02)&(0.11)&(0.12)&&(0.50)&(0.27)&&\multicolumn{1}{c}{(...)}&\multicolumn{1}{c}{(...)}&\multicolumn{1}{c}{(...)}&\multicolumn{1}{c}{(...)} \\
J1243$+$4646& 21.46& 21.48& 21.63& 20.33& 21.14&& 21.31& 21.61&&\multicolumn{1}{c}{ ... }&\multicolumn{1}{c}{ ... }&\multicolumn{1}{c}{ ... }&\multicolumn{1}{c}{ ... } \\
            &(0.11)&(0.04)&(0.07)&(0.03)&(0.22)&&(0.26)&(0.24)&&\multicolumn{1}{c}{(...)}&\multicolumn{1}{c}{(...)}&\multicolumn{1}{c}{(...)}&\multicolumn{1}{c}{(...)} \\
J1248$+$4259& 21.55& 21.22& 20.94& 21.23& 20.65&& 21.42& 21.16&&\multicolumn{1}{c}{ ... }&\multicolumn{1}{c}{ ... }&\multicolumn{1}{c}{ ... }&\multicolumn{1}{c}{ ... } \\
            &(0.13)&(0.05)&(0.05)&(0.09)&(0.20)&&(0.38)&(0.28)&&\multicolumn{1}{c}{(...)}&\multicolumn{1}{c}{(...)}&\multicolumn{1}{c}{(...)}&\multicolumn{1}{c}{(...)} \\
J1256$+$4509& 22.35& 22.01& 21.43& 22.18& 21.51&& 21.66& 21.85&& 16.29& 15.80& 12.94&\multicolumn{1}{c}{ ... } \\
            &(0.18)&(0.06)&(0.05)&(0.15)&(0.26)&&(0.30)&(0.27)&&\multicolumn{1}{c}{(0.06)}&\multicolumn{1}{c}{(0.11)}&\multicolumn{1}{c}{(0.54)}&\multicolumn{1}{c}{(...)} \\
\hline
\end{tabular}

  \end{table*}

\begin{figure*}
\hbox{
\includegraphics[angle=0,width=0.19\linewidth]{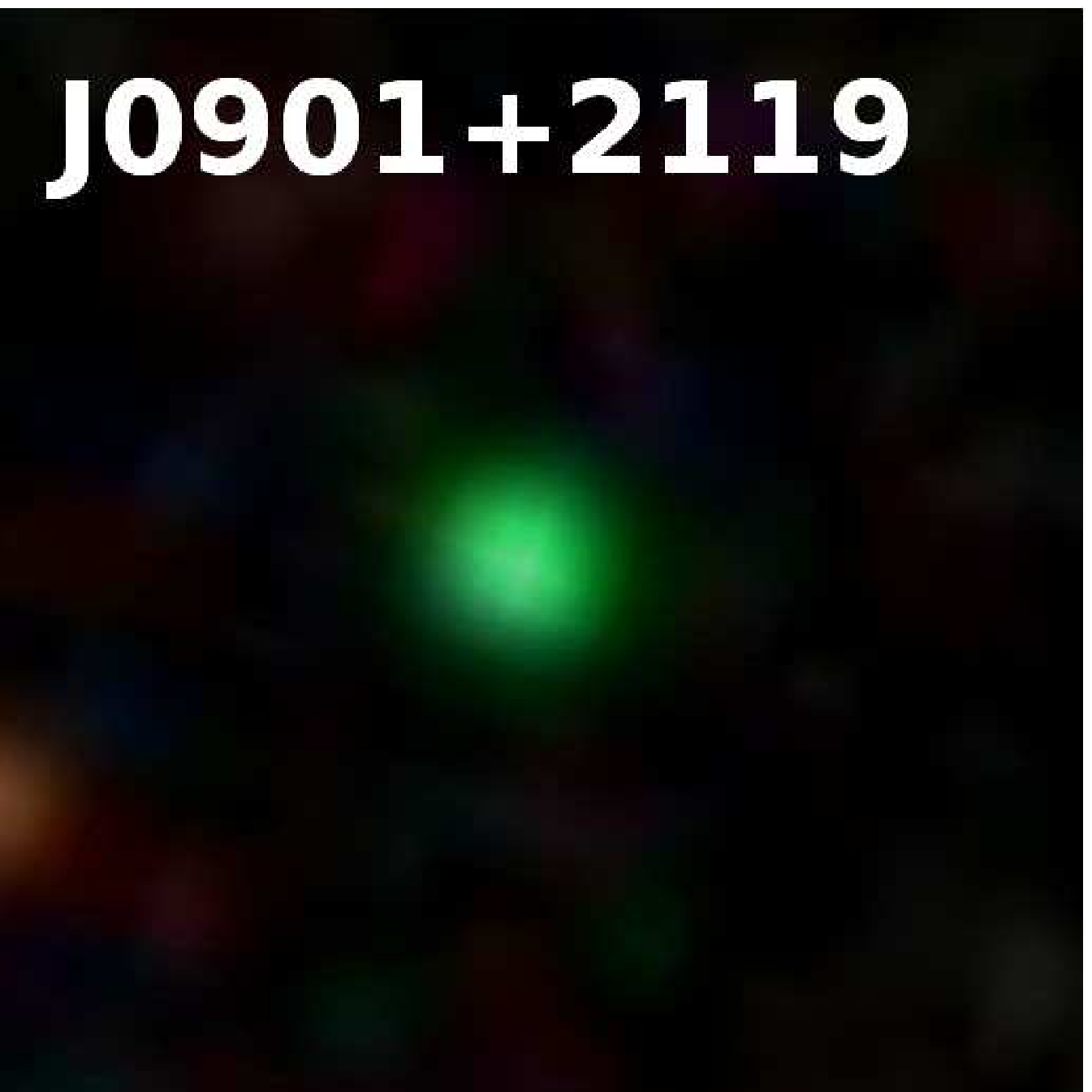}
\includegraphics[angle=0,width=0.19\linewidth]{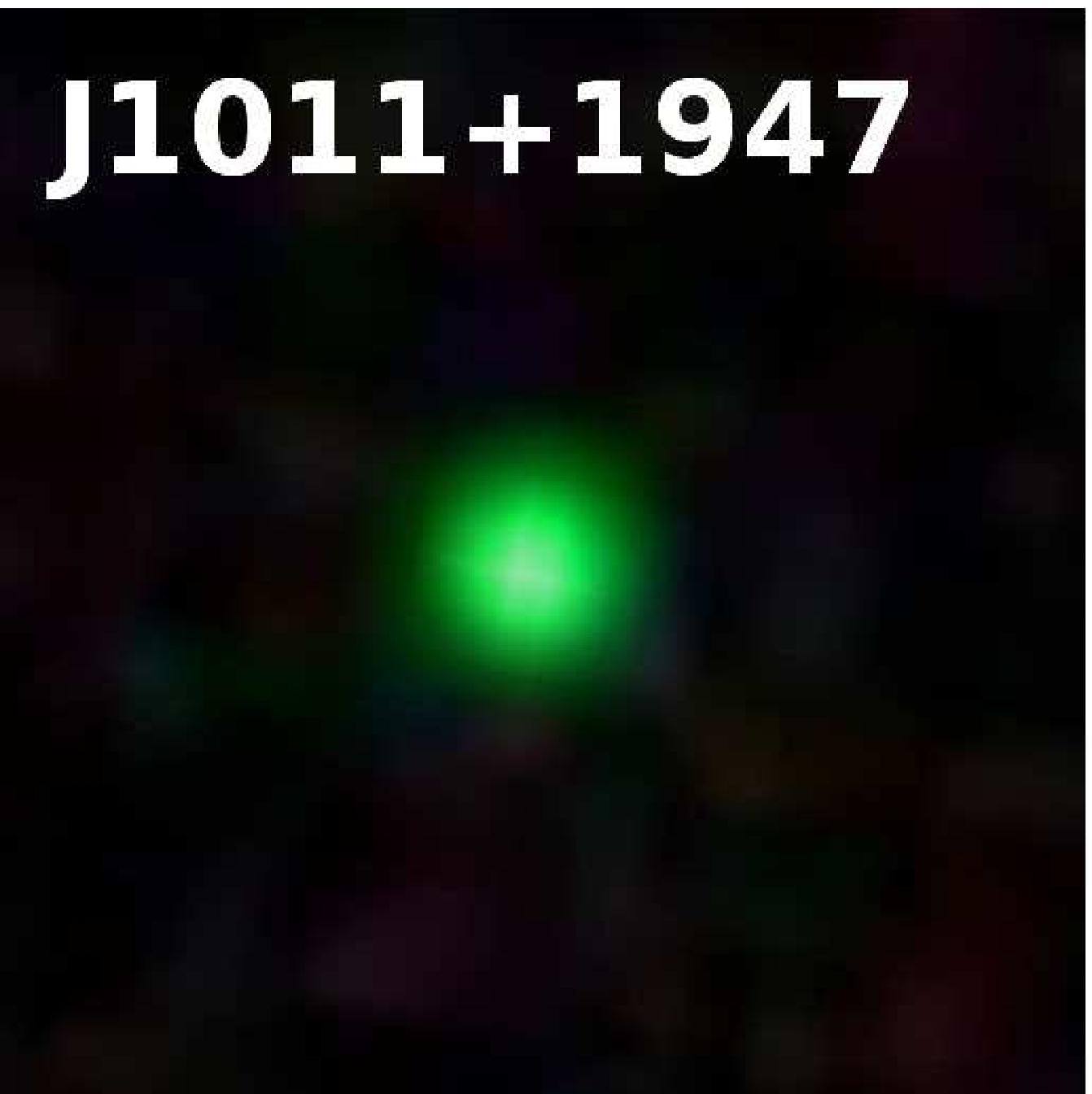}
\includegraphics[angle=0,width=0.19\linewidth]{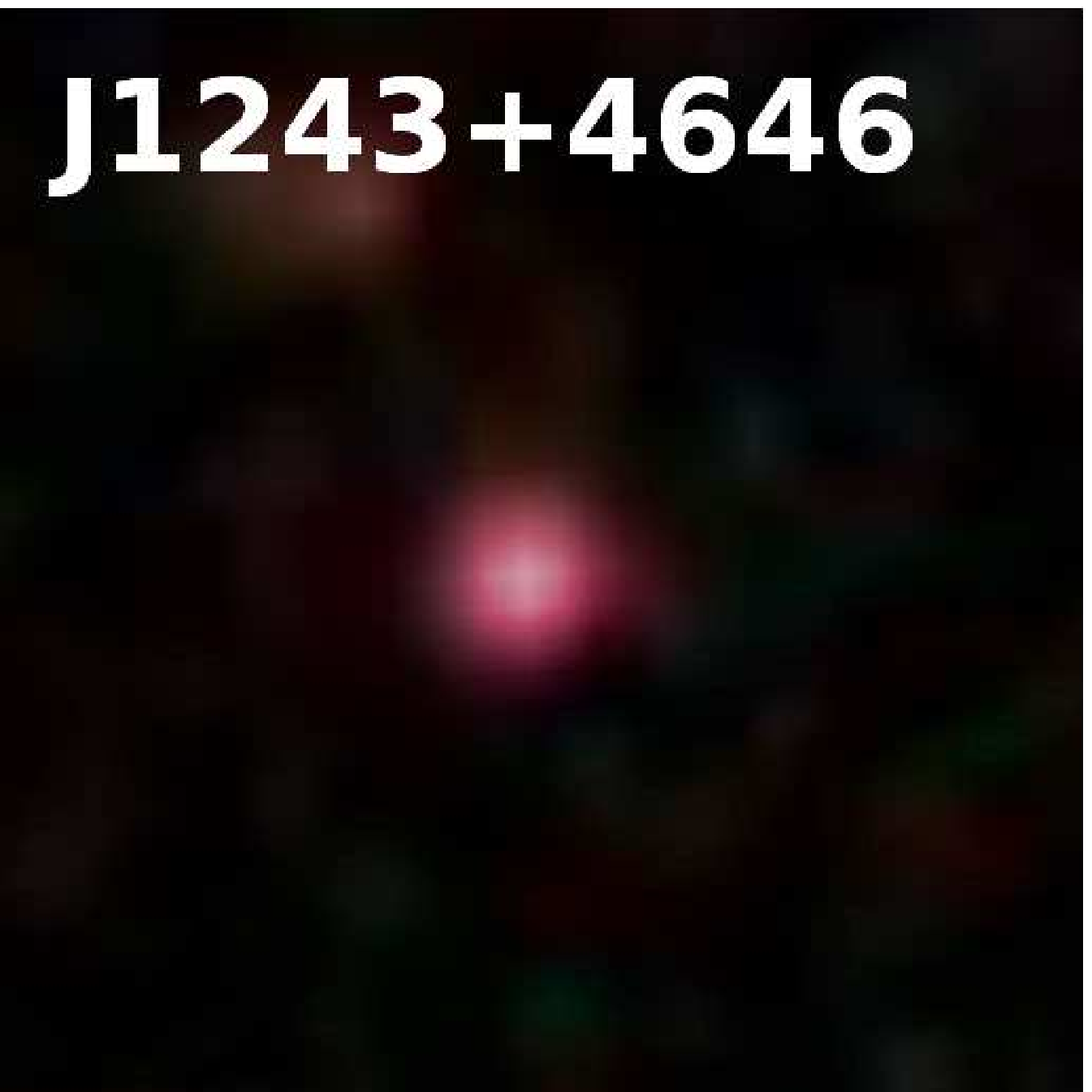}
\includegraphics[angle=0,width=0.19\linewidth]{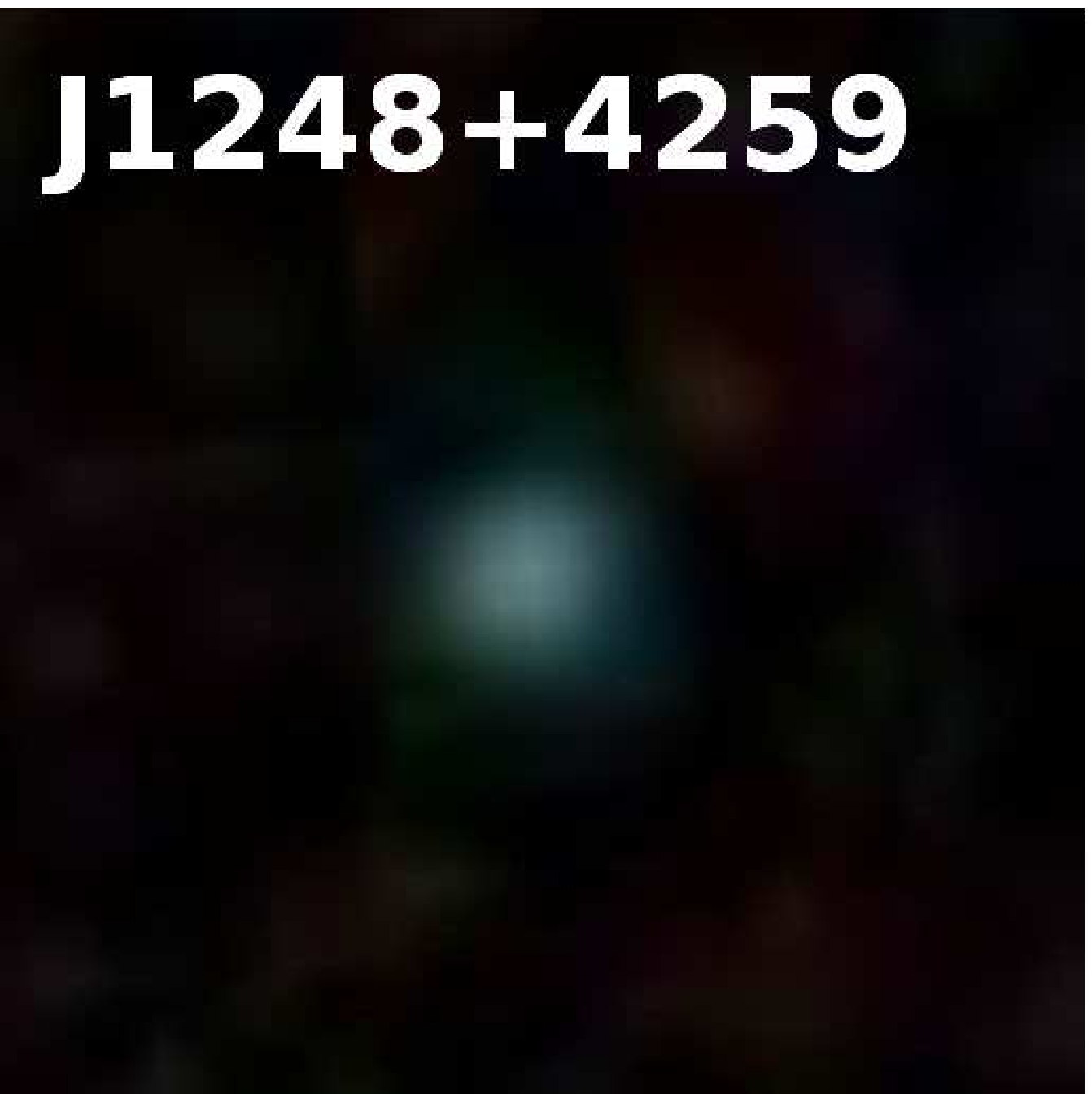}
\includegraphics[angle=0,width=0.19\linewidth]{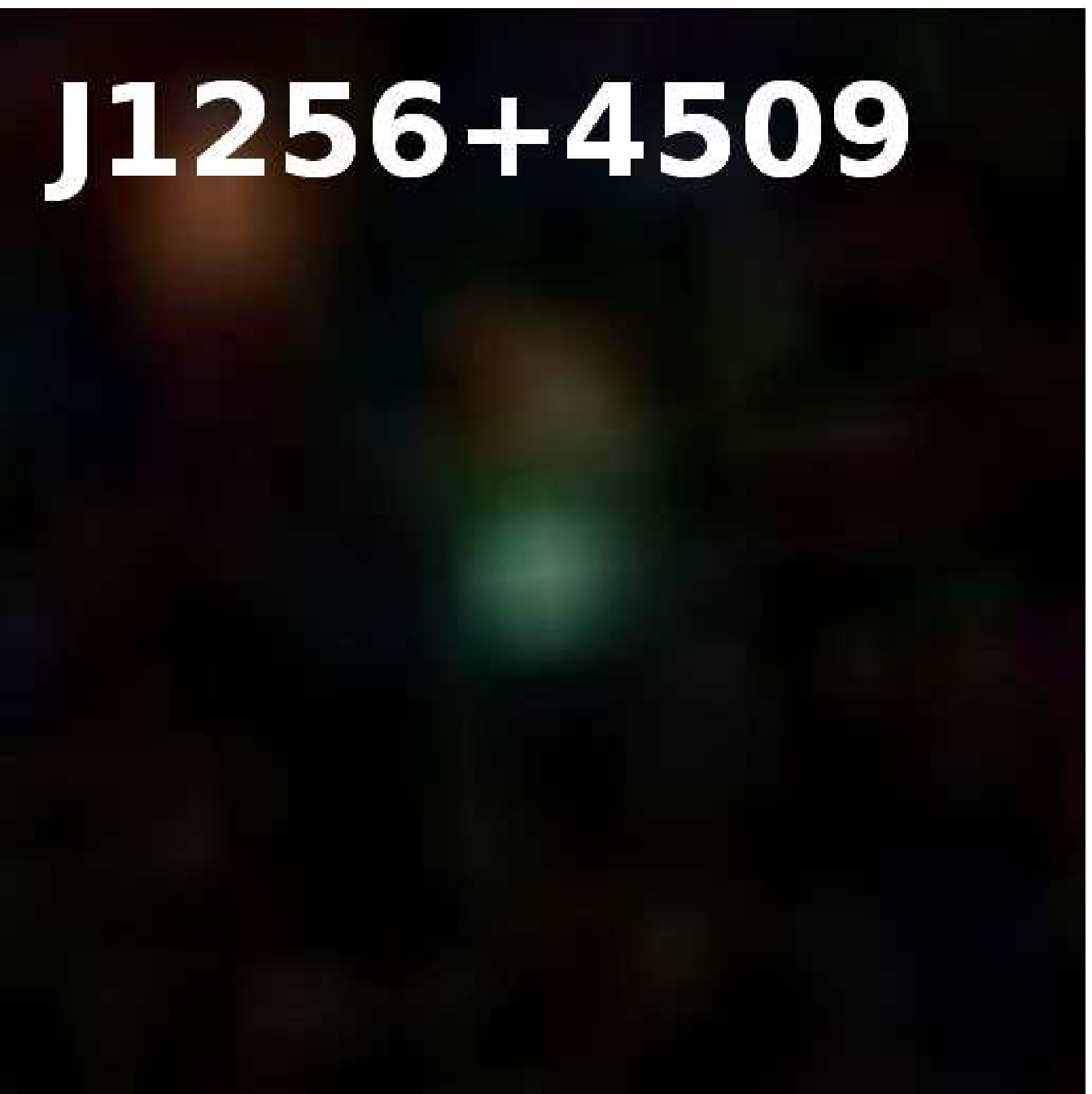}
}
\caption{25 arcsec $\times$ 25 arcsec SDSS composite images of selected galaxies.
\label{fig1}}
\end{figure*}

\begin{figure*}
\includegraphics[angle=-90,width=0.49\linewidth]{diagnDR7_c_1.ps}
\includegraphics[angle=-90,width=0.49\linewidth]{oiii_oii_c.ps}
\caption{{\bf a)} The Baldwin-Phillips-Terlevich (BPT) diagram \citep{BPT81} for
SFGs. {\bf b)} The O$_{32}$ -- R$_{23}$ diagram for SFGs where R$_{23}$ = 
([O~{\sc ii}]3727 + [O~{\sc iii}]4959 + [O~{\sc iii}]5007)/H$\beta$.
The LyC leaking galaxies (this paper) and  J1154$+$2443 \citep{I18}
are shown by filled stars in both panels.
The location of LyC leaking galaxies by \citet{B14,C17} are shown by open
circles while LyC leaking galaxies by \citet{I16,I16b} are represented by
open stars. The compact SFGs \citep{I16c} are represented by grey dots.
The solid line in {\bf a)} \citep{K03} separates SFGs from active galactic 
nuclei.
\label{fig2}}
\end{figure*}

In this paper we present new {\sl HST}/COS observations of the LyC 
in 5 compact SFGs with the highest O$_{32}$ $\sim$ 8 -- 27
ever observed in order to detect ionizing radiation, and examine its
behaviour over a wide range of O$_{32}$. 
The properties of selected SFGs derived from observations in the optical range
are presented in Section \ref{sec:select}. The {\sl HST} observations 
and data reduction are described
in Section \ref{sec:obs}. The surface brightness profiles 
in the UV range are discussed in Section~\ref{sec:sbp}. 
In Section \ref{sec:global} we compare the {\sl HST}/COS spectra
with the extrapolation of the modelled SEDs to the UV range.
Ly$\alpha$ emission is considered in 
Section \ref{sec:lya}. The escaping Lyman continuum emission is discussed in 
Section \ref{sec:lyc} together with the corresponding
escape fractions. In Section \ref{discussion} 
our results are compared with the LyC escape fractions
for other galaxies obtained in some recent studies.
We summarize our findings in Section \ref{summary}.

\section{Properties of selected galaxies derived from
observations in the optical range}\label{sec:select}

\subsection{Emission-line diagnostic diagrams}

Our LyC leaker candidates were selected from the SDSS Data Release 12 (DR12) 
\citep{A15} by adopting the selection criteria described in 
\citet{I16,I16b,I18}. All of them have a
compact structure on SDSS images (Fig.~\ref{fig1}) and high equivalent 
widths EW(H$\beta$) $>$ 200~\AA\
of the H$\beta$ emission line in the SDSS spectra, 
indicating very recent star formation.
These galaxies are located in the upper part of the SFG branch in the 
Baldwin-Phillips-Terlevich (BPT) diagram (filled stars in 
Fig. \ref{fig2}a) implying the presence of high-excitation H~{\sc ii} regions.
\citet{I16,I16b} discussed only galaxies 
with O$_{32}$ $\sim$ 5 -- 7. In this paper and in \citet{I18} 
we report observations of galaxies with the considerably larger range of 
O$_{32}$ $\sim$ 8 -- 27 (Table \ref{tab1}), aiming to study the applicability 
of the O$_{32}$ criterion to select galaxies with high escape fractions of 
ionizing radiation. It is seen in Fig.~\ref{fig2}b that 
the selected galaxies have the highest O$_{32}$ ratios among all local LyC 
leaking galaxies observed so far.

The SDSS, {\sl GALEX} and {\sl WISE} apparent magnitudes of the selected
galaxies are shown in Table \ref{tab2}, indicating that these SFGs are
among the faintest low-redshift LyC leakers that have been observed 
with {\sl HST}.

  \begin{table*}
  \caption{Extinction-corrected flux densities and rest-frame equivalent widths of
the emission lines in SDSS spectra
\label{tab3}}
\begin{tabular}{lcrrrrrrrrrr} \hline
 & &\multicolumn{10}{c}{Galaxy}\\
Line &\multicolumn{1}{c}{$\lambda$}&\multicolumn{2}{c}{J0901$+$2119}& \multicolumn{2}{c}{J1011$+$1947}& \multicolumn{2}{c}{J1243$+$4646}& \multicolumn{2}{c}{J1248$+$4259}& \multicolumn{2}{c}{J1256$+$4509} \\
     &&\multicolumn{1}{c}{$I$$^{\rm a}$}&\multicolumn{1}{c}{EW$^{\rm b}$}&\multicolumn{1}{c}{$I$$^{\rm a}$}&\multicolumn{1}{c}{EW$^{\rm b}$}&\multicolumn{1}{c}{$I$$^{\rm a}$}&\multicolumn{1}{c}{EW$^{\rm b}$}&\multicolumn{1}{c}{$I$$^{\rm a}$}&\multicolumn{1}{c}{EW$^{\rm b}$}&\multicolumn{1}{c}{$I$$^{\rm a}$}&\multicolumn{1}{c}{EW$^{\rm b}$} \\
\hline
Mg~{\sc ii}          &2796&  \multicolumn{1}{c}{...}& ...&\multicolumn{1}{c}{...}& ...&\multicolumn{1}{c}{...}& ...&  10.8$\pm$2.4&  12&  17.2$\pm$4.2&   8\\
Mg~{\sc ii}          &2803&  \multicolumn{1}{c}{...}& ...&\multicolumn{1}{c}{...}& ...&\multicolumn{1}{c}{...}& ...&   6.5$\pm$2.3&   7&   8.6$\pm$3.8&   4\\
$[$O~{\sc ii}$]$     &3727&  81.5$\pm$4.0& 196& 29.7$\pm$2.3&  64& 53.7$\pm$4.5&  81& 49.4$\pm$3.1& 137& 44.3$\pm$4.4&  92\\
H12                  &3750&  \multicolumn{1}{c}{...}& ...&\multicolumn{1}{c}{...}& ...&\multicolumn{1}{c}{...}& ...&   3.3$\pm$1.9&   9&   4.0$\pm$2.9&   6\\
H11                  &3771&  \multicolumn{1}{c}{...}& ...&\multicolumn{1}{c}{...}& ...&  5.6$\pm$2.8&  13&   5.1$\pm$1.9&  13&   6.5$\pm$2.9&   9\\
H10                  &3798&  \multicolumn{1}{c}{...}& ...&\multicolumn{1}{c}{...}& ...&  7.9$\pm$3.3&  17&   8.0$\pm$1.9&  21&   4.7$\pm$2.8&  12\\
H9                   &3836&  \multicolumn{1}{c}{...}& ...& 6.2$\pm$2.9&  28&  8.9$\pm$3.4&  19&   9.4$\pm$1.9&  28&   7.1$\pm$3.3&  14\\
$[$Ne~{\sc iii}$]$   &3869&  46.4$\pm$2.9&  97& 49.7$\pm$3.0& 103& 49.1$\pm$4.3&  98& 46.3$\pm$2.9& 151& 55.1$\pm$4.7& 120\\
H8+He~{\sc i}        &3889&  13.5$\pm$3.2&  33& 14.6$\pm$2.8&  34& 20.5$\pm$3.8&  38& 20.4$\pm$2.2&  87& 19.5$\pm$3.4&  41\\
H7+$[$Ne~{\sc iii}$]$&3969&  21.6$\pm$3.4&  52& 31.8$\pm$2.9&  96& 39.1$\pm$4.1&  80& 33.9$\pm$2.6& 124& 33.7$\pm$4.0&  69\\
H$\delta$            &4101&  24.2$\pm$3.6&  56& 26.3$\pm$3.6&  47& 26.4$\pm$3.4& 106& 28.1$\pm$2.4& 112& 26.7$\pm$3.8&  41\\
H$\gamma$            &4340&  48.0$\pm$4.4&  97& 44.5$\pm$3.6&  98& 47.2$\pm$4.2& 134& 49.6$\pm$2.9& 228& 48.2$\pm$4.5&  93\\
$[$O~{\sc iii}$]$    &4363&   7.6$\pm$1.3&  19& 14.6$\pm$1.7&  29& 15.5$\pm$2.8&  45& 17.7$\pm$1.9&  96& 16.4$\pm$2.9&  39\\
He~{\sc i}           &4471&   3.9$\pm$1.0&  14&  2.8$\pm$0.9&   8&  3.1$\pm$2.1&   9&  4.2$\pm$1.3&  18&  4.2$\pm$2.3&   8\\
H$\beta$             &4861& 100.0$\pm$4.9& 255&100.0$\pm$4.8& 237&100.0$\pm$5.7& 221&100.0$\pm$4.0& 426&100.0$\pm$5.9& 253\\
$[$O~{\sc iii}$]$    &4959& 221.2$\pm$6.9& 741&261.9$\pm$7.9& 731&240.9$\pm$8.9& 738&194.0$\pm$6.1& 851&249.6$\pm$9.8& 654\\
$[$O~{\sc iii}$]$    &5007& 654.6$\pm$26.&1481&807.2$\pm$19.&2304&725.9$\pm$20.&2057&583.6$\pm$14.&2138&723.3$\pm$21.&1872\\
He~{\sc i}           &5876&   9.6$\pm$1.2&  35& 11.7$\pm$1.3& 103&\multicolumn{1}{c}{...}& ...& 10.8$\pm$1.3&  81& 11.4$\pm$2.1&  52\\
$[$O~{\sc i}$]$      &6300&   3.2$\pm$0.9&  10&  2.2$\pm$0.6&  28&\multicolumn{1}{c}{...}& ...&  2.2$\pm$0.7&  49&  \multicolumn{1}{c}{...}\\
$[$S~{\sc iii}$]$    &6312&  \multicolumn{1}{c}{...}& ...&  4.3$\pm$0.8&  29&\multicolumn{1}{c}{...}& ...&  \multicolumn{1}{c}{...}& ...&  \multicolumn{1}{c}{...}& ...\\
H$\alpha$            &6563& 288.8$\pm$9.0& 831&283.7$\pm$8.9&1052&280.9$\pm$10.& 740&279.8$\pm$8.2&2561&280.5$\pm$11.& 955\\
$[$N~{\sc ii}$]$     &6583&   9.7$\pm$1.3&  22&  7.1$\pm$1.1&  22&  5.8$\pm$1.8&  10&  5.7$\pm$0.9&  58&  7.9$\pm$1.2&  34\\
He~{\sc i}           &6678&  \multicolumn{1}{c}{...}& ...&\multicolumn{1}{c}{...}& ...&\multicolumn{1}{c}{...}& ...&  2.6$\pm$0.4&  30&  \multicolumn{1}{c}{...}& ...\\
$[$S~{\sc ii}$]$     &6717&   5.5$\pm$0.9&  57&\multicolumn{1}{c}{...}& ...&\multicolumn{1}{c}{...}& ...&  2.6$\pm$0.7&  32&  3.8$\pm$1.3&  26\\
$[$S~{\sc ii}$]$     &6731&   5.4$\pm$0.9&  40&\multicolumn{1}{c}{...}& ...&\multicolumn{1}{c}{...}& ...&  3.1$\pm$0.7&  40&  5.6$\pm$1.5&  38\\
He~{\sc i}           &7065&  \multicolumn{1}{c}{...}& ...&\multicolumn{1}{c}{...}& ...&\multicolumn{1}{c}{...}& ...&  5.1$\pm$0.8&  58&  \multicolumn{1}{c}{...}& ...\\
$[$Ar~{\sc iii}$]$   &7136&  \multicolumn{1}{c}{...}& ...&\multicolumn{1}{c}{...}& ...&\multicolumn{1}{c}{...}& ...&  3.7$\pm$0.7&  72&  \multicolumn{1}{c}{...}& ...\\
$C$(H$\beta$)$_{\rm int}$$^{\rm c}$  &&\multicolumn{2}{c}{0.185$\pm$0.036}&\multicolumn{2}{c}{0.135$\pm$0.037}&\multicolumn{2}{c}{0.090$\pm$0.044}&\multicolumn{2}{c}{0.185$\pm$0.035}&\multicolumn{2}{c}{0.095$\pm$0.047}\\
$C$(H$\beta$)$_{\rm MW}$$^{\rm d}$   &&\multicolumn{2}{c}{0.038}&\multicolumn{2}{c}{0.037}&\multicolumn{2}{c}{0.017}&\multicolumn{2}{c}{0.032}&\multicolumn{2}{c}{0.027}\\
EW(H$\beta$)$^{\rm b}$        &&\multicolumn{2}{c}{255$\pm$16}&\multicolumn{2}{c}{237$\pm$30}&\multicolumn{2}{c}{221$\pm$10}&\multicolumn{2}{c}{426$\pm$12}&\multicolumn{2}{c}{253$\pm$17}\\
$I$(H$\beta$)$^{\rm e}$     &&\multicolumn{2}{c}{29.1$\pm$1.4}&\multicolumn{2}{c}{27.0$\pm$1.3}&\multicolumn{2}{c}{14.1$\pm$0.8}&\multicolumn{2}{c}{35.2$\pm$1.4}&\multicolumn{2}{c}{11.4$\pm$0.7}\\
\hline
  \end{tabular}

\hbox{$^{\rm a}$$I$=100$\times$$I$($\lambda$)/$I$(H$\beta$) where $I$($\lambda$) 
and $I$(H$\beta$) are flux densities of emission lines, corrected for both the 
Milky Way and internal}

\hbox{~\,extinction.}

\hbox{$^{\rm b}$Rest-frame equivalent width in \AA.}

\hbox{$^{\rm c}$Internal galaxy extinction coefficient.}

\hbox{$^{\rm d}$Milky Way extinction coefficient.}


\hbox{$^{\rm e}$in 10$^{-16}$ erg s$^{-1}$ cm$^{-2}$.}

  \end{table*}

  \begin{table*}
  \caption{Electron temperatures, electron number densities and 
element abundances in H~{\sc ii} regions \label{tab4}}
  \begin{tabular}{lccccc} \hline
Galaxy &J0901$+$2119 &J1011$+$1947 &J1243$+$4646 &J1248$+$4259 &J1256$+$4509  \\ \hline
$T_{\rm e}$ ($[$O {\sc iii}$]$), K      & 12190$\pm$780       & 14610$\pm$750       & 15690$\pm$1350     & 18820$\pm$1160& 16060$\pm$1410        \\
$T_{\rm e}$ ($[$O {\sc ii}$]$), K       & 12000$\pm$720       & 13770$\pm$660       & 14370$\pm$1160     & 15460$\pm$890 & 14550$\pm$1190        \\
$T_{\rm e}$ ($[$S {\sc iii}$]$), K      & 12170$\pm$650       & 13280$\pm$620       & 14360$\pm$1120     & 17580$\pm$970 & 14720$\pm$1170        \\
$N_{\rm e}$ ($[$S {\sc ii}$]$), cm$^{-3}$&   530$\pm$600       &100$^{\rm a}$  &        100$^{\rm a}$     &   1180$\pm$1180  &   2690$\pm$2690        \\ \\
O$^+$/H$^+$$\times$10$^{5}$             &1.67$\pm$0.18        &0.35$\pm$0.03        &0.56$\pm$0.07       &0.46$\pm$0.03 &0.57$\pm$0.06          \\
O$^{2+}$/H$^+$$\times$10$^{5}$          &12.72$\pm$0.68        &9.42$\pm$0.25        &7.13$\pm$0.21      &3.79$\pm$0.08  &6.78$\pm$0.08           \\
O/H$\times$10$^{5}$                    &14.39$\pm$0.70        &9.78$\pm$0.26        &7.68$\pm$0.23      &4.37$\pm$0.11  &7.35$\pm$0.11          \\
12+log O/H                             &8.16$\pm$0.02        &7.99$\pm$0.01        &7.89$\pm$0.01      &7.64$\pm$0.01  &7.87$\pm$0.01          \\ \\
N$^+$/H$^+$$\times$10$^{6}$             &1.15$\pm$0.16        &0.62$\pm$0.10        &0.47$\pm$0.10      &0.40$\pm$0.07  &0.64$\pm$0.15          \\
ICF(N)$^{\rm b}$                        &7.55                 &22.48                 &12.02               &8.69         &11.25                   \\
N/H$\times$10$^{6}$                     &8.69$\pm$1.37       &13.95$\pm$2.85        &5.60$\pm$2.06      &3.50$\pm$0.64 &7.16$\pm$1.89          \\
log N/O                                &~$-$1.22$\pm$0.07~~\, &~$-$0.85$\pm$0.09~~\, &~$-$1.14$\pm$0.16~~\,&~$-$1.07$\pm$0.12~~\, &~$-$1.01$\pm$0.12~~\,      \\ \\
Ne$^{2+}$/H$^+$$\times$10$^{5}$          &2.41$\pm$0.29        &1.44$\pm$0.12        &1.15$\pm$0.13       &0.68$\pm$0.05 &1.21$\pm$0.14          \\
ICF(Ne)$^{\rm b}$                       &1.03                 &0.97                 &1.01                &1.06           &1.01                   \\
Ne/H$\times$10$^{5}$                    &2.47$\pm$0.29        &1.39$\pm$0.11        &1.17$\pm$0.14       &0.71$\pm$0.05 &1.23$\pm$0.14          \\
log Ne/O                               &~$-$0.75$\pm$0.12~~\, &~$-$0.85$\pm$0.04~~\, &~$-$0.82$\pm$0.05~~\,&~$-$0.79$\pm$0.03~~\, &~$-$0.78$\pm$0.05~~\,    \\ \\
Mg$^{+}$/H$^+$$\times$10$^{6}$          &      ...            &  ...               &  ...              &0.10$\pm$0.02 &0.20$\pm$0.05          \\
ICF(Mg)$^{\rm b}$                        &      ...            &  ...               &  ...              &13.73        &21.03                  \\
Mg/H$\times$10$^{6}$                    &      ...            &  ...               &  ...              &1.39$\pm$0.29 &4.17$\pm$1.04         \\
log Mg/O                               &      ...            &  ...               &  ...              &~$-$1.50$\pm$0.09~~\, &~$-$1.25$\pm$0.11~~\,   \\ \\
Ar$^{2+}$/H$^+$$\times$10$^{7}$          &      ...            &  ...               &  ...              &1.17$\pm$0.24  &  ...                     \\
ICF(Ar)$^{\rm b}$                        &      ...            &  ...               &  ...              &1.34           &  ...                     \\
Ar/H$\times$10$^{7}$                    &      ...            &  ...               &  ...              &1.58$\pm$0.95  &  ...                   \\
log Ar/O                               &      ...            &  ...               &  ...              &~$-$2.44$\pm$0.26~~\, &  ...              \\ 
\hline
\end{tabular}

\hbox{$^{\rm a}$Assumed value.}
\hbox{$^{\rm b}$Ionization correction factor.}
  \end{table*}

\subsection{Interstellar extinction and element abundances}\label{sec:ext}

To derive interstellar extinction and ionized gas metallicity we use the SDSS 
spectra of the selected galaxies. Our approach is described in detail in 
\citet{I16,I16b,I18}. 
In short, we use the prescriptions by \citet*{ITL94}
to derive galaxy internal interstellar extinction from the observed decrements
of hydrogen emission lines. Then the extinction-corrected emission lines are
used to derive ionic and total element abundances following the methods described in
\citet{I06} and \citet{G13}.

The emission-line flux densities in the observed SDSS spectra, 
uncorrected for redshift, were first corrected for 
the Milky Way extinction with $A(V)_{\rm MW}$ from the NASA Extragalactic 
Database (NED)\footnote{NASA/IPAC Extragalactic Database (NED) is operated by 
the Jet Propulsion Laboratory, California Institute of Technology, under 
contract with the National Aeronautics and Space Administration.}, adopting 
\citet*{C89}' reddening law and $R(V)_{\rm MW} = 3.1$.
Then, the flux densities of emission lines at the rest-frame 
wavelengths were corrected for the internal 
extinction of galaxies with $R(V)_{\rm int} = 3.1$ and $A(V)_{\rm int}$ =
3.1$\times$$E(B-V)_{\rm int}$, where $E(B-V)_{\rm int}$ = 
$C$(H$\beta$)$_{\rm int}$/1.47 \citep{A84}.

The emission-line flux densities corrected for both the Milky Way
and internal extinction, the rest-frame equivalent widths, 
the Milky Way ($C$(H$\beta$)$_{\rm MW}$) and internal 
($C$(H$\beta$)$_{\rm int}$) extinction coefficients,
and extinction-corrected H$\beta$ flux densities are shown in Table \ref{tab3}.
%
The flux densities from Table \ref{tab3} and the direct $T_{\rm e}$ method
are used to derive 
physical conditions (the electron temperature and electron number density) and 
the element abundances in the H~{\sc ii} regions.
These quantities are shown in Table \ref{tab4}. The oxygen abundances are 
comparable to those in known low-redshift LyC leakers 
\citep{I16,I16b,I18}. The ratios of the neon, magnesium and argon
abundances to oxygen abundance are similar to those in
dwarf emission-line galaxies \citep[e.g. ][]{I06,G13}. On the other hand, the 
nitrogen-to-oxygen abundance ratios are somewhat elevated, similar to those in
other LyC leakers at $z$~$\ga$~0.3.

\subsection{Luminosities and stellar masses}

The emission-line luminosities and stellar masses of our galaxies were obtained 
adopting a luminosity distance \citep[NED,][]{W06} with the cosmological 
parameters $H_0$=67.1 km s$^{-1}$Mpc$^{-1}$, $\Omega_\Lambda$=0.682, 
$\Omega_m$=0.318 \citep{P14}.

The H$\beta$ luminosity $L$(H$\beta$) and corresponding
star-formation rates SFR were obtained from the extinction-corrected H$\beta$ 
flux densities using the relation by \citet{K98} for SFR. 
The star-formation rate should be increased by a factor 1/[1-$f_{\rm esc}$(LyC)], 
to take into account the escaping ionizing radiation which is discussed later
(see Sect.~\ref{sec:lyc}). The SFRs corrected for escaping LyC radiation are
shown in Table \ref{tab5}, and they are in the range of values for other LyC leakers.
Their specific star formation rates
sSFR = SFR/$M_\star$ are also similar to sSFRs for other LyC leakers 
\citep{I16,I16b,I18} and are among the highest known for dwarf SFGs
at any redshift \citep{I16c}.


We use SDSS spectra of our LyC leakers to fit the spectral energy 
distribution (SED) and to derive their stellar masses. 
The fitting method, using a two-component model, is described in \citet{I18}. To take into account
the contribution of the young stellar population we adopt a single
instantaneous burst. For the older stellar population, we assume that it
was formed continuously with a constant SFR. 
Since the H$\beta$ equivalent widths of our SFGs are very high, nebular 
continuum emission must be taken into account; it is determined from the 
observed H$\beta$ recombination line 
flux density and knowing the ISM temperature and density.
A $\chi ^2$ minimization technique was used to find the best model for
the continuum. An additional requirement is that the modelled H$\beta$
and H$\alpha$ equivalent widths should reproduce the observed values.
Several parameters (the starburst age, the age of the older stellar 
population, the young-to-old stellar population mass ratio) were varied 
with a Monte-Carlo method to determine the best fit.

\begin{figure*}
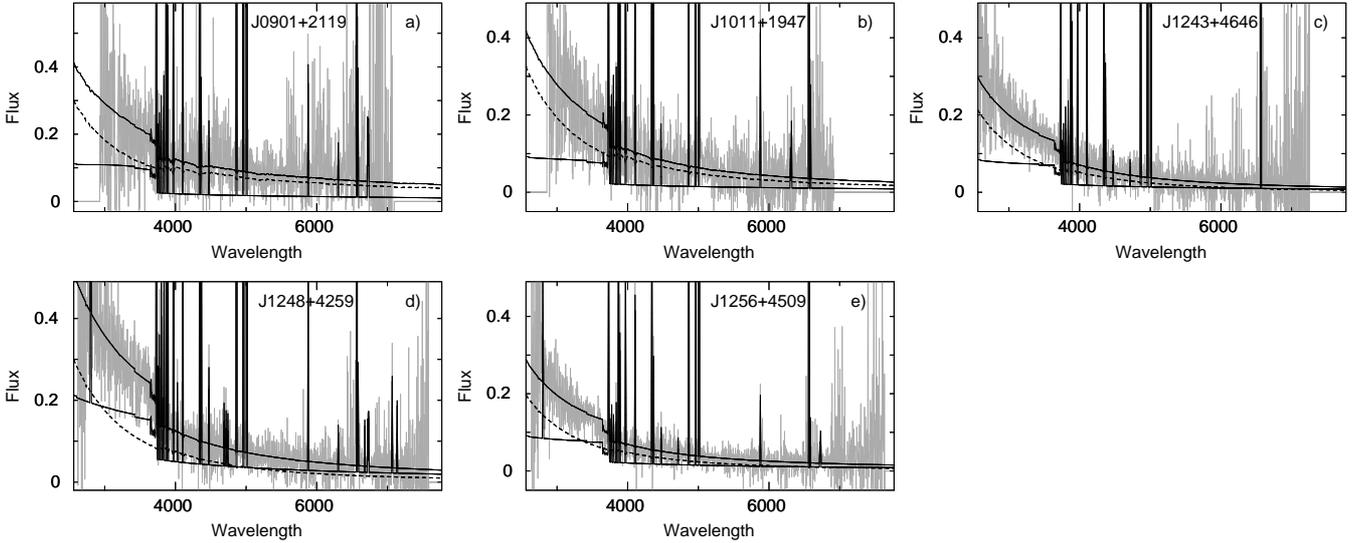

\hbox{
\includegraphics[angle=-90,width=0.33\linewidth]{mainsp_J0901+2119_c_1.ps}
\includegraphics[angle=-90,width=0.33\linewidth]{mainsp_J1011+1947_c_1.ps}
%
\includegraphics[angle=-90,width=0.33\linewidth]{mainsp_J1243+4646_c_1.ps}
 }
 \hbox{
\includegraphics[angle=-90,width=0.33\linewidth]{mainsp_J1248+4259_c_1.ps}
%
\includegraphics[angle=-90,width=0.33\linewidth]{mainsp_J1256+4509_c_1.ps}
}
\caption{SED fitting of the galaxy SDSS spectra. 
The rest-frame extinction-corrected spectra are shown by grey
lines. The total, nebular and stellar modelled SEDs are shown by thick solid, 
thin solid and dashed lines, respectively. Flux densities and wavelengths are
expressed in 10$^{-16}$ erg s$^{-1}$ cm$^{-2}$ \AA$^{-1}$ and \AA, respectively.
\label{fig3}}
\end{figure*}

To illustrate the quality of our SED fitting we show in Fig. \ref{fig3}
the modelled stellar, nebular and total SEDs superposed to the rest-frame
extinction-corrected SDSS spectra. For all galaxies we find very good 
agreement. We note the considerable contribution ($\sim$ 20 -- 50 per cent)
of nebular emission to the total (stellar+nebular) continuum because of
the very high equivalent widths of hydrogen emission lines indicating
that the emission in the optical range is mostly produced by very
young starbursts.

The total stellar masses (young plus old population) and starburst ages (of the young population) 
of our LyC leakers derived from SED fitting are presented in Table \ref{tab5}. They are similar to those 
derived for other LyC leakers \citep{I16,I16b,I18}. The mass of the young
stellar population (not shown in Table \ref{tab5}) is typically a small fraction ($\sim 1-3$ per cent)  of the total stellar mass
in all our sources.

  \begin{table*}
  \caption{Integrated characteristics \label{tab5}}
  \begin{tabular}{lccccccccccc} \hline
Name&$M_{\rm FUV}$$^{\rm a}$&  $M_g$$^{\rm b}$   &log $M_\star$$^{\rm c}$ &SB age&log $L$(H$\beta$)$^{\rm d}$&SFR$^{\rm e}$&log sSFR$^{\rm e}$&$\alpha$$^{\rm f}$&$r_{50}$$^{\rm g}$&$\Sigma$$^{\rm h}$&$\Sigma$$^{\rm i}$\\
    & (mag)  & (mag)  &(log M$_\odot$)& (Myr)&(log erg s$^{-1}$)&(M$_\odot$ yr$^{-1}$)&(log yr$^{-1}$)&(kpc)&(kpc)& 
\multicolumn{2}{c}{(M$_\odot$ yr$^{-1}$kpc$^{-2}$)}  \\ \hline   
J0901$+$2119&$-$20.02&$-$19.53&9.8&2.4&41.93&20&$-$8.5&1.51&0.18&~\,2.7&192\\
J1011$+$1947&$-$20.24&$-$20.07&9.0&3.4&42.00&25&$-$7.6&1.48&0.15&~\,4.1&397\\
J1243$+$4646&$-$20.49&$-$20.46&7.8&2.7&41.99&80&$-$5.9&1.75&0.19&~\,8.4&719\\
J1248$+$4259&$-$20.60&$-$20.32&8.2&2.5&42.21&37&$-$6.6&1.64&0.22&~\,4.3&242\\
J1256$+$4509&$-$19.95&$-$19.45&8.8&2.4&41.69&18&$-$7.6&1.72&0.17&~\,2.0&202\\
\hline
  \end{tabular}

\hbox{$^{\rm a}$Absolute FUV magnitude derived from the intrinsic rest-frame SED.}

\hbox{$^{\rm b}$Absolute SDSS $g$ magnitude corrected for the Milky Way extinction.}

\hbox{$^{\rm c}$$M_\star$ is the total stellar mass (young $+$ older population).}

\hbox{$^{\rm d}$$L$(H$\beta$) is the H$\beta$ luminosity corrected for the Milky Way and internal extinction.}

\hbox{$^{\rm e}$Corrected for the Milky Way and internal extinction, and escaping LyC radiation.}

\hbox{$^{\rm f}$Exponential disc scale length.}

\hbox{$^{\rm g}$Galaxy radius where NUV intensity equal to half of maximal intensity.}

\hbox{$^{\rm h}$Star-formation rate surface density assuming galaxy radius equal to $\alpha$.}

\hbox{$^{\rm i}$Star-formation rate surface density assuming galaxy radius equal to $r_{50}$.}

  \end{table*}

  \begin{table}
  \caption{{\sl HST}/COS observations \label{tab6}}
  \begin{tabular}{lcccc} \hline
\multicolumn{1}{c}{}&\multicolumn{1}{c}{}&\multicolumn{3}{c}{Exposure time (s)} \\ 
\multicolumn{1}{c}{Name}&\multicolumn{1}{c}{Date}&\multicolumn{3}{c}{(Central wavelength (\AA))} \\ 
    &    &MIRRORA&G140L&G160M \\ \hline
J0901$+$2119&2017-12-26&2$\times$1408     &  5636& 5636\\
            &          &         &(1105)&(1623)\\
J1011$+$1947&2017-12-22&2$\times$1408     &  5753& 5637\\
            &          &         &(1105)&(1577)\\
J1243$+$4646&2017-12-19&2$\times$1473     &  9081& 5897\\
            &          &         &(1105)&(1611)\\
J1248$+$4259&2017-12-02&2$\times$1451     &  5809& 5809\\
            &          &         &(1105)&(1600)\\ 
J1256$+$4509&2017-12-16&2$\times$1473     &  9082& 5896\\
            &          &         &(1105)&(1600)\\ 
\hline
\end{tabular}
  \end{table}

\begin{figure*}
\hbox{
\includegraphics[angle=0,width=0.33\linewidth]{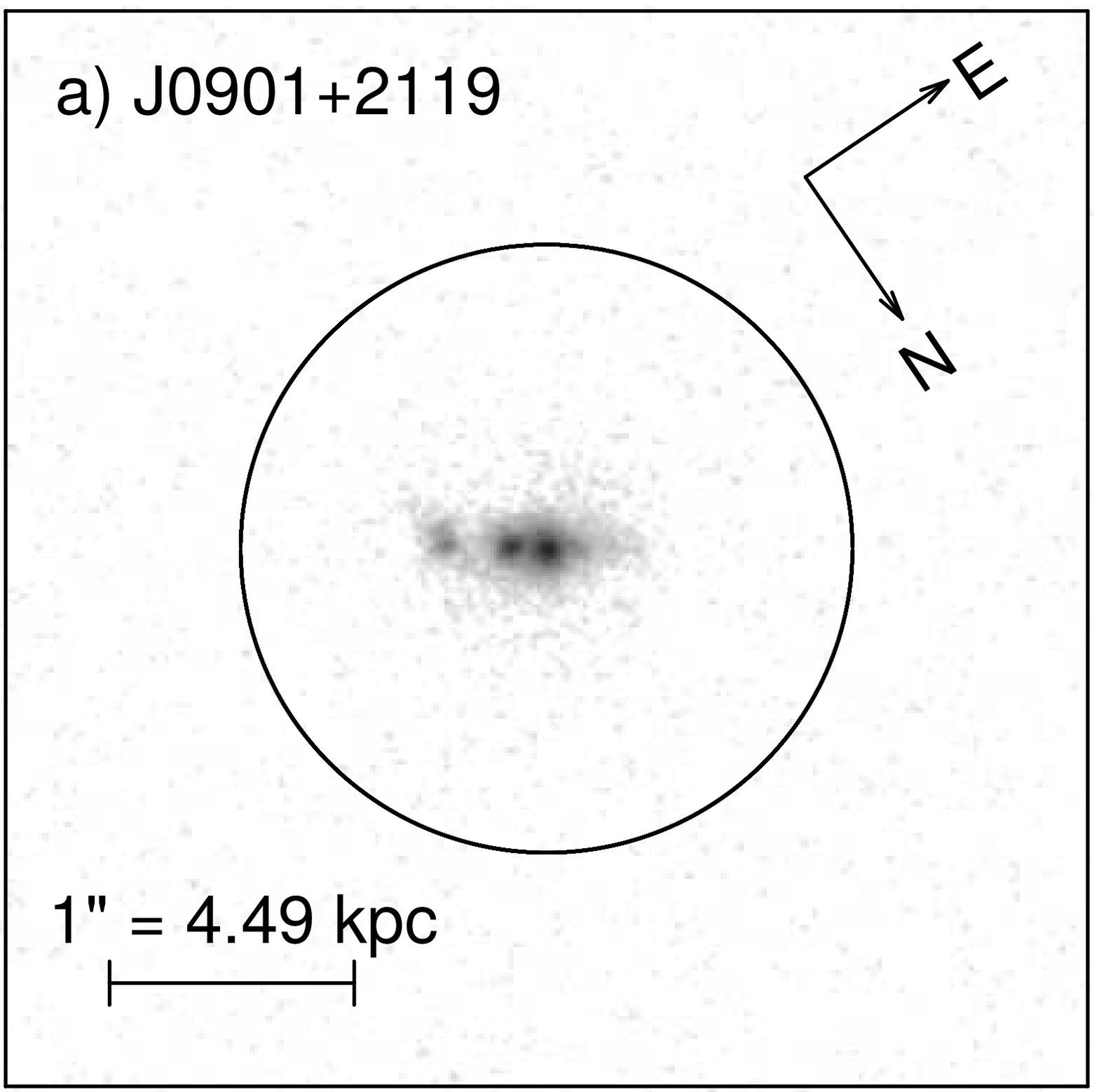}
\includegraphics[angle=0,width=0.33\linewidth]{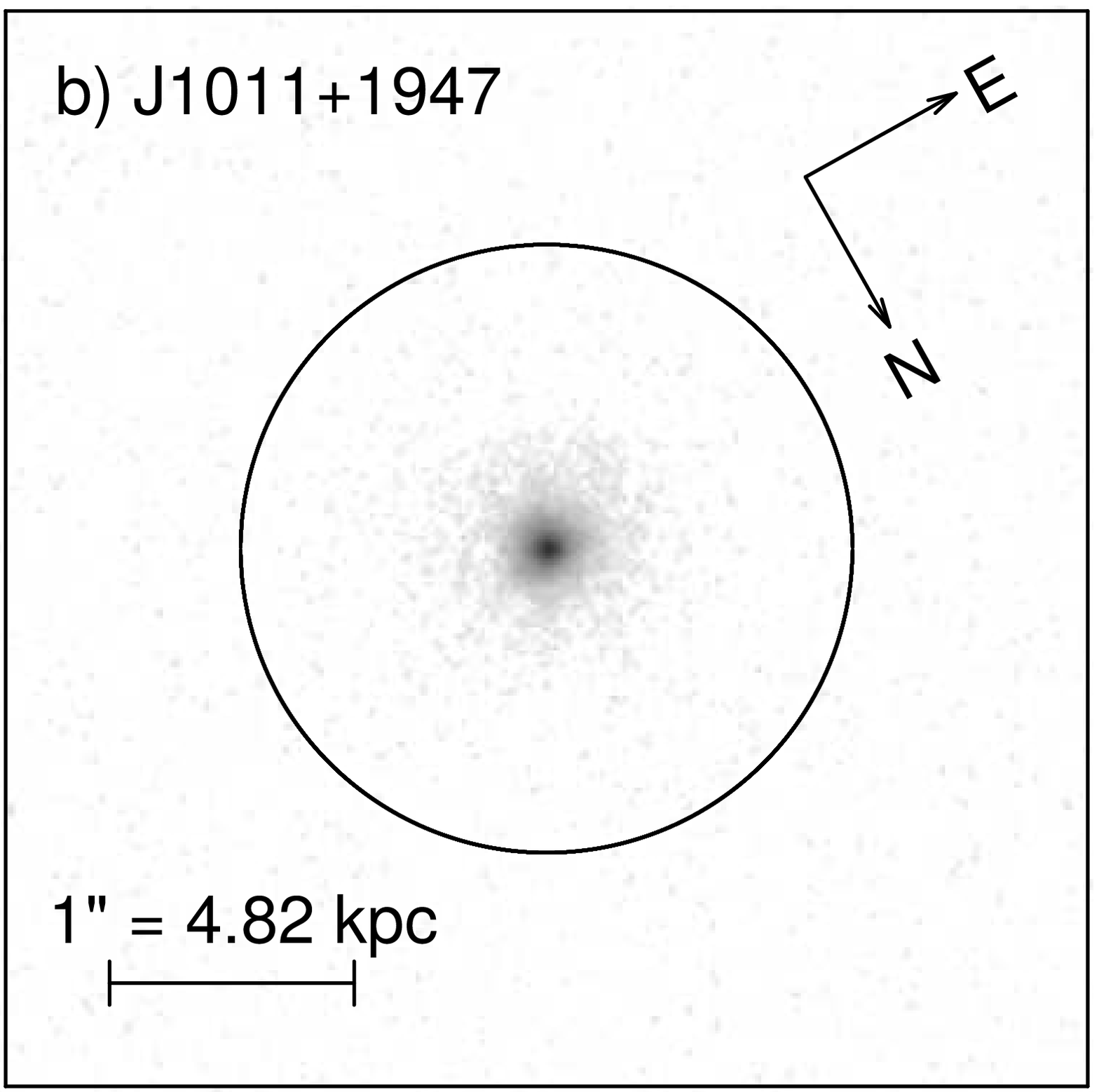}
\includegraphics[angle=0,width=0.33\linewidth]{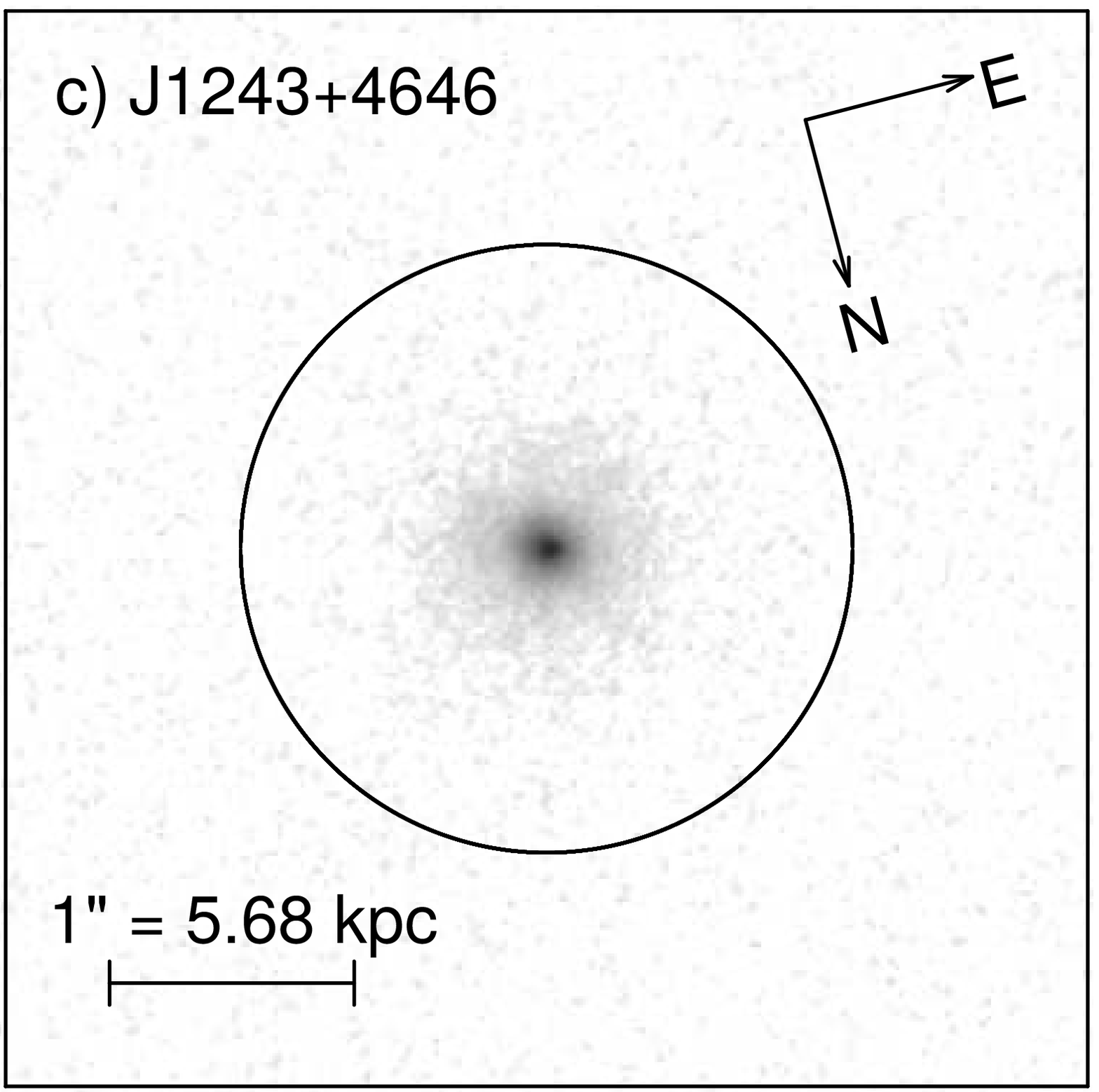}
}
\hbox{
\includegraphics[angle=0,width=0.33\linewidth]{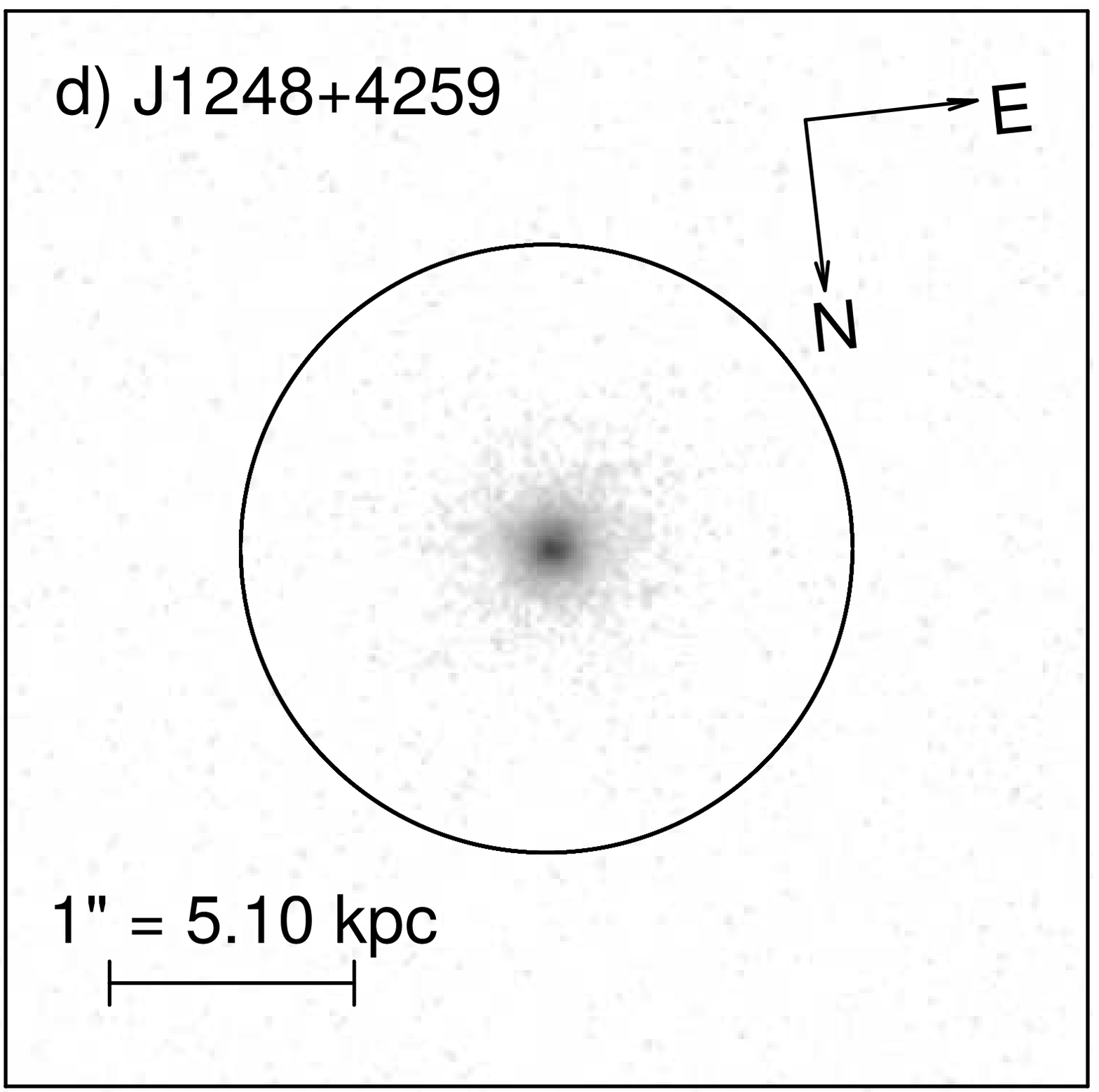}
\includegraphics[angle=0,width=0.33\linewidth]{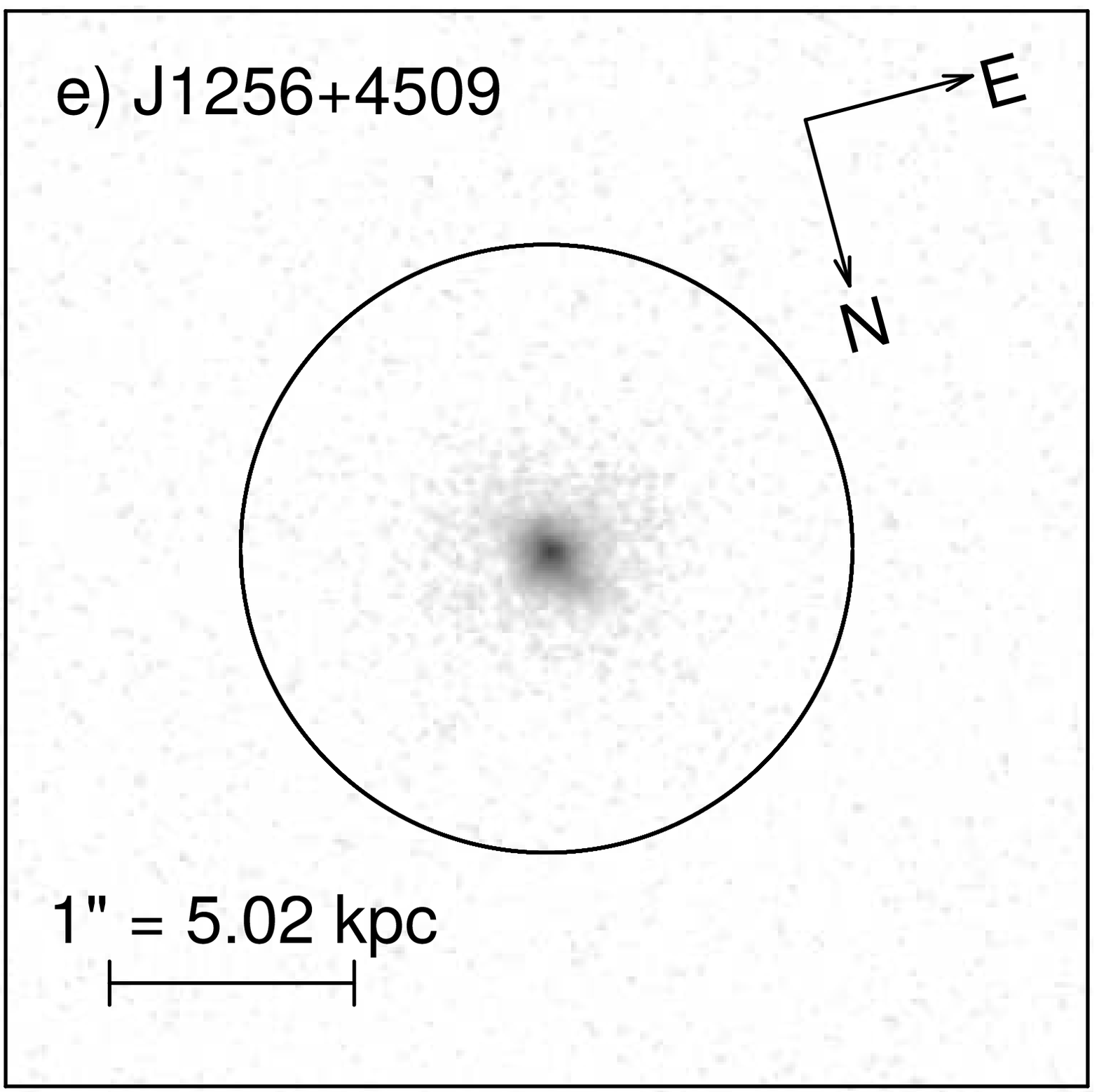}
}
\caption{The {\sl HST} NUV acquisition images of the LyC leaking galaxies in 
log surface brightness scale. The COS spectroscopic aperture with a diameter of 
2.5 arcsec is shown in all panels by a circle.
The linear scale in each panel is derived adopting an angular size distance.
\label{fig4}}
\end{figure*}

\section{{\sl HST}/COS observations and data 
reduction}\label{sec:obs}

{\sl HST}/COS spectroscopy of the five LyC leaker candidates was obtained
in program GO~14635 (PI: Y.\ I.\ Izotov) in December 2017.
The program also included J1154$+$2443, the data of which was
already published \citep{I18}. The observational details
are presented in Table \ref{tab6}. As in our previous programs
\citep{I16,I16b}, the galaxies were acquired by COS near
ultraviolet (NUV) imaging. Since our targets are compact but faint, as
based on shallow {\sl GALEX} imaging, one entire orbit per object was spent 
for deep NUV imaging and reliable acquisition.
The NUV-brightest region of each target was centered in the
$\sim 2.5$\,arcsec diameter spectroscopic aperture (Fig. \ref{fig4}).
Although the galaxies show some structure with an extended
low-surface-brightness (LSB) component and, in the case of
J0901$+$2119, several star-forming knots, their sizes are smaller
than the central unvignetted $0.8$\,arcsec diameter region of
the spectroscopic aperture \citep{F18}. Hence, the galaxy quantities 
derived from the COS spectra do not require corrections for vignetting.

The spectra were obtained at COS Lifetime Position 4 with the 
low-resolution grating G140L and medium-resolution grating G160M, applying all 
four focal-plane offset positions.
One orbit of G140L observations of J1011$+$1947 failed due to a guide
star re-acquisition error, shortening the total exposure to 5753\,s.
However, the shorter exposure was sufficient to achieve our science
goals. The 1105\,\AA\ setup was used for the G140L grating 
(COS Lifetime Position 4: wavelength range 1110--2150\,\AA, resolving
power $R\simeq 1400$ at 1150\,\AA) to include the redshifted LyC
emission for all targets. We obtained resolved spectra of the galaxies'
Ly$\alpha$ emission lines with the G160M grating ($R\sim 16000$ at
1600\,\AA), varying the G160M central wavelength with galaxy redshift
to cover the emission line and the nearby continuum on one detector
segment.

The individual exposures were reduced with the \textsc{CALCOS} pipeline v3.2.1,
followed by accurate background subtraction and co-addition with custom
software \citep{W16}. We used the same methods and extraction
aperture sizes as in \citet{I18} to achieve a homogeneous
reduction of the galaxy sample observed in our program. We checked the
accuracy of our custom correction for scattered light in COS G140L data
by comparing the LyC flux densities obtained in the total exposure and in 
orbital night, respectively (see below). Particularly in the LyC range of the
targets the estimated relative background error of 4--10 per cent does
not significantly affect our results.

\begin{figure*}
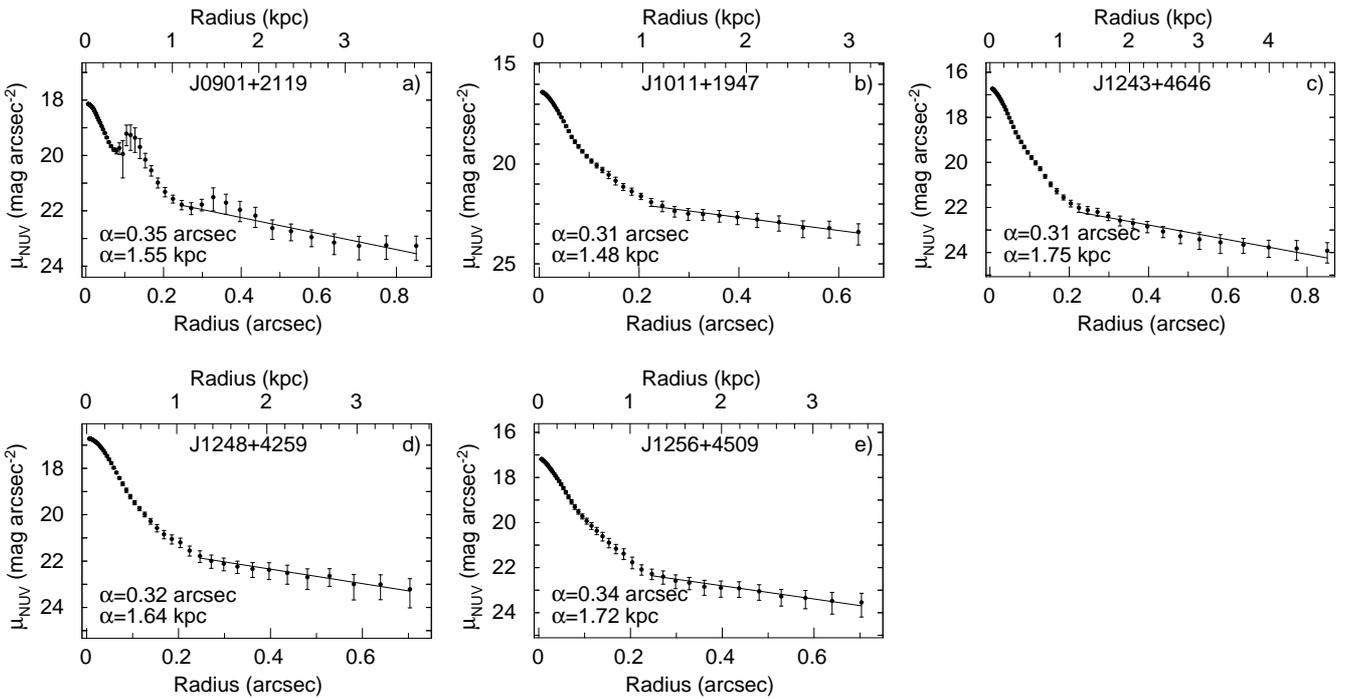

\hbox{
\includegraphics[angle=-90,width=0.33\linewidth]{SBP_J0901+2119.ps}
\includegraphics[angle=-90,width=0.33\linewidth]{SBP_J1011+1947.ps}
%
\includegraphics[angle=-90,width=0.33\linewidth]{SBP_J1243+4646.ps}
 }
 \hbox{
\includegraphics[angle=-90,width=0.33\linewidth]{SBP_J1248+4259.ps}
%
\includegraphics[angle=-90,width=0.33\linewidth]{SBP_J1256+4509.ps}
}
\caption{NUV surface brightness profiles of galaxies. The linear fits are 
shown for the range of radii used for fitting.
\label{fig5}}
\end{figure*}

\section{Surface brightness distribution in the NUV range}\label{sec:sbp}

To determine the surface brightess (SB) profiles of our galaxies we use the COS NUV 
acquisition images. The method uses the routine {\it ellipse} in 
{\sc iraf}\footnote{{\sc iraf} is distributed by 
the National Optical Astronomy Observatories, which are operated by the 
Association of Universities for Research in Astronomy, Inc., under cooperative 
agreement with the National Science Foundation.}/{\sc stsdas}\footnote{{\sc stsdas} is a product of 
the Space Telescope Science Institute, which is operated by AURA for NASA.}
and is described in detail e.g. in \citet{I18}. 
The SB profiles (Fig. \ref{fig5}) are common
to all LyC leakers studied thus far \citep[this paper, ][]{I16,I16b,I18}
with a sharp increase in the central part because of the presence
of the bright star-forming region(s) and 
a linear decrease (in magnitudes) in the outer part, 
reminiscent of a disc structure.

The scale lengths $\alpha$ of our galaxies defined in Eq.~1 of  \citet{I16b}
are in the range $\sim$ 1.5 -- 1.8 kpc (Fig.~\ref{fig5}), somewhat higher than 
$\alpha$ = 0.6 -- 1.4 kpc in other LyC leakers \citep{I16,I16b,I18}. The 
corresponding surface densities of star-formation rate in the studied galaxies 
$\Sigma$ = SFR/($\pi \alpha^2$) are
somewhat lower than in other LyC leakers, mainly because of their higher
$\alpha$. Because of the compactness of the bright star-forming region, the 
half-light radii $r_{50}$ of our galaxies in the NUV are 
considerably smaller than $\alpha$ (see Table \ref{tab5}).
Adopting $r_{50}$ as a measure of the size of these galaxies, 
the corresponding $\Sigma$s are
typically two orders of magnitude larger, and comparable to those found for 
SFGs at high redshifts \citep{CL16,PA18,Bo17}. We note for completeness that the
star-formation rate surface densities in the LyC leaking galaxy J1154$+$2443,
observed in the course of the same {\sl HST} GO 14635 program, are
5.1 M$_\odot$ yr$^{-1}$ kpc$^{-2}$ and 150 M$_\odot$ yr$^{-1}$ kpc$^{-2}$ for
$\alpha$ = 1.09 kpc and $r_{50}$ = 0.2 kpc, respectively \citep{I18}, 
similar to the respective values for the SFGs shown in Table \ref{tab5}.

\section{Comparison of the {\sl HST}/COS spectra with the modelled
SEDs in the UV range}\label{sec:global}



\begin{figure*}
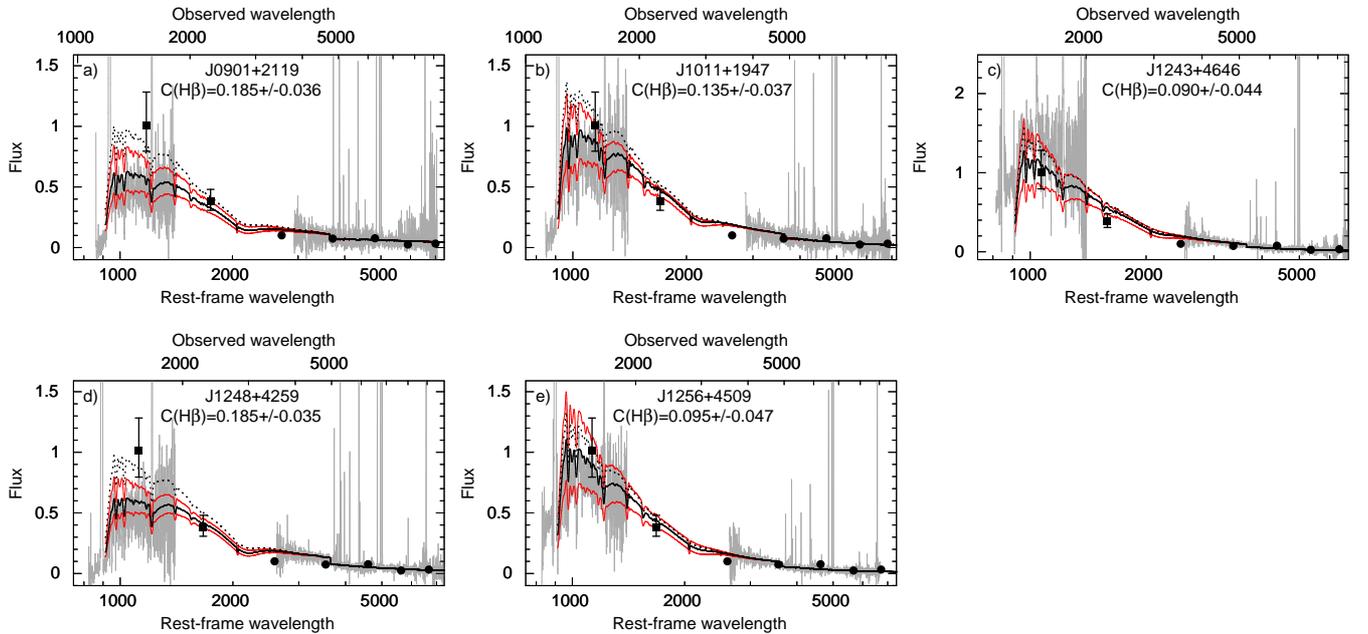

\hbox{
\includegraphics[angle=-90,width=0.33\linewidth]{spectrum_tot1_c_j0901+2119.ps}
\includegraphics[angle=-90,width=0.33\linewidth]{spectrum_tot1_c_j1011+1947.ps}
%
\includegraphics[angle=-90,width=0.33\linewidth]{spectrum_tot1_c_j1243+4646.ps}
 }
 \hbox{
\includegraphics[angle=-90,width=0.33\linewidth]{spectrum_tot1_c_j1248+4259.ps}
%
\includegraphics[angle=-90,width=0.33\linewidth]{spectrum_tot1_c_j1256+4509.ps}
}
\caption{A comparison of the COS G140L and SDSS spectra (grey lines), and
photometric data (filled squares and filled circles) with the modelled SEDs.
Modelled SEDs reddened by the Milky Way 
extinction with $R(V)_{\rm MW}$ = 3.1 and internal extinction with $R(V)_{\rm int}$
= 3.1 and 2.4 are shown by thick
dotted and solid lines, respectively. The SED variations with the extinction
coefficient $C$(H$\beta$) in the 1$\sigma$ range and $R(V)_{\rm int}$ = 2.4 are 
shown by red thin lines. Flux densities are in 
10$^{-16}$ erg s$^{-1}$ cm$^{-2}$\AA$^{-1}$, wavelengths are in \AA. \label{fig6}}
\end{figure*}

To derive the fraction of the escaping ionizing radiation, one 
of the two methods which we use \citep{I18} is based on the comparison between 
the observed flux density in the Lyman continuum range and the intrinsic 
flux density produced by stellar populations in the galaxy. 
The latter can be obtained from SED fitting of the SDSS spectra.
To verify the quality of our SED fitting, we extrapolate the reddened SEDs 
to the UV range and compare them with the observed COS spectra in 
Fig. \ref{fig6}. For comparison, we also show by filled symbols the {\sl GALEX} 
FUV and NUV flux densities and the flux densities in the SDSS $u,g,r,i,z$ 
filters. We find that the spectroscopic 
and photometric data in the optical range are consistent, indicating that almost
all the emission of our galaxies is inside the spectroscopic aperture. 
Therefore, aperture corrections are not needed. It can be noted that the 
{\sl GALEX} FUV photometric flux densities deviate somewhat from the 
spectroscopic COS flux densities with the two most deviant cases
being J0901$+$2119 and J1248$+$4259. The main reason for these deviations
is the non-negligible contribution of the redshifted Ly$\alpha$ emission 
line to the {\sl GALEX} FUV band. 
Indeed, a \lya\ emission line with an equivalent width in the range
$\sim$ 90 -- 250\AA\ (see Table \ref{tab7}) may contribute $\sim$ 20 -- 50 
per cent to the emission in the {\sl GALEX} FUV band with an effective width
of 255\AA\ \citep*{Rod13}.
However, even for J0901$+$2119 and J1248$+$4259 which 
show the highest \lya\ equivalent widths, the FUV 
photometric and spectroscopic data are consistent within 2$\sigma$ errors. 

We show in Fig. \ref{fig6} 
the modelled intrinsic SEDs reddened by adopting the extinction 
coefficients $C$(H$\beta$)$_{\rm MW}$ and $C$(H$\beta$)$_{\rm int}$ 
(Table \ref{tab3}) and the reddening law by 
\citet{C89} with $R(V)_{\rm MW}$ = 3.1 and $R(V)_{\rm int}$ = 2.4 (thick solid 
lines) and 3.1 (dotted lines). For $R(V)_{\rm int}$ = 2.4, we show by red
thin solid lines the variations of the reddened SEDs produced by 1$\sigma$
variations of $C$(H$\beta$)$_{\rm int}$.

It is seen in Fig. \ref{fig6} that the models reproduce the SDSS 
spectra quite well and do not depend on the adopted $R(V)_{\rm int}$ and 
variations within 1$\sigma$ uncertainties of $C$(H$\beta$)$_{\rm int}$. However, 
in the UV range there is a stronger dependence of the reddened SEDs on both 
$R(V)_{\rm int}$ and $C$(H$\beta$)$_{\rm int}$ variations. The modelled SEDs 
reddened with $R(V)_{\rm int}$ = 2.4 reproduce best the COS G140L 
spectra, in agreement with the conclusions reached by \citet{I16,I16b,I18} for 
the other LyC leakers. This implies that the reddening law in the UV range is 
steeper in our SFGs than in the Milky Way. An exception is J1243$+$4646, for 
which the reddening law with $R(V)_{\rm int}$ = 3.1 works better, in apparent
contradiction with other SFGs. However, an unusual
Ly$\alpha$ profile with three peaks (Sect. \ref{sec:lya}) 
and a very high $f_{\rm esc}$(LyC) (Sect. \ref{sec:lyc}) imply that
the interstellar medium in J1243$+$4646 is clumpy with dust-free holes
allowing for some UV continuum radiation to escape through these holes.
This could mimick a flatter attenuation law \citep[see eq.~2 in ][]{G18}
as compared to that with $R(V)_{\rm int}$ = 2.4.

  \begin{table*}
  \caption{Parameters for the Ly$\alpha$ emission line \label{tab7}}
  \begin{tabular}{rcrcrr} \hline
Name&$A$(Ly$\alpha$)$_{\rm MW}$$^{\rm a}$&\multicolumn{1}{c}{$I$$^{\rm b}$}&log $L$$^{\rm c}$&\multicolumn{1}{c}{EW$^{\rm d}$}&\multicolumn{1}{c}{$V_{\rm sep}$$^{\rm e}$} \\ 
\hline
J0901$+$2119&0.210& 106.2$\pm$2.3&42.49&179$\pm$3.9& 345.0$\pm$12.5\\
J1011$+$1947&0.202& 123.4$\pm$2.6&42.66&121$\pm$3.1& 276.4$\pm$\,~5.4\\
J1243$+$4646&0.092& 180.3$\pm$2.8&43.09& 98$\pm$1.9& 143.4$\pm$\,~4.0\\
            &     &              &     &           & 163.8$\pm$\,~5.7\\
J1248$+$4259&0.174& 150.0$\pm$2.7&42.83&256$\pm$5.2& 283.8$\pm$15.9\\
J1256$+$4509&0.147&  88.4$\pm$2.1&42.58& 86$\pm$3.2& 239.4$\pm$10.5\\
\hline
  \end{tabular}

\hbox{$^{\rm a}$Milky Way extinction at the observed wavelength of the Ly$\alpha$
emission line in mags}

\hbox{\, adopting \citet{C89} reddening law with $R(V)$=3.1.}

\hbox{$^{\rm b}$Flux density in 10$^{-16}$ erg s$^{-1}$ cm$^{-2}$ measured in
the COS spectrum and corrected for the Milky Way extinction.}

\hbox{$^{\rm c}$$L$ is Ly$\alpha$ luminosity in erg s$^{-1}$ corrected for the
Milky Way extinction.}

\hbox{$^{\rm d}$Rest-frame equivalent width in \AA.}

\hbox{$^{\rm e}$Ly$\alpha$ peak separation in km s$^{-1}$.}

  \end{table*}

\begin{figure*}
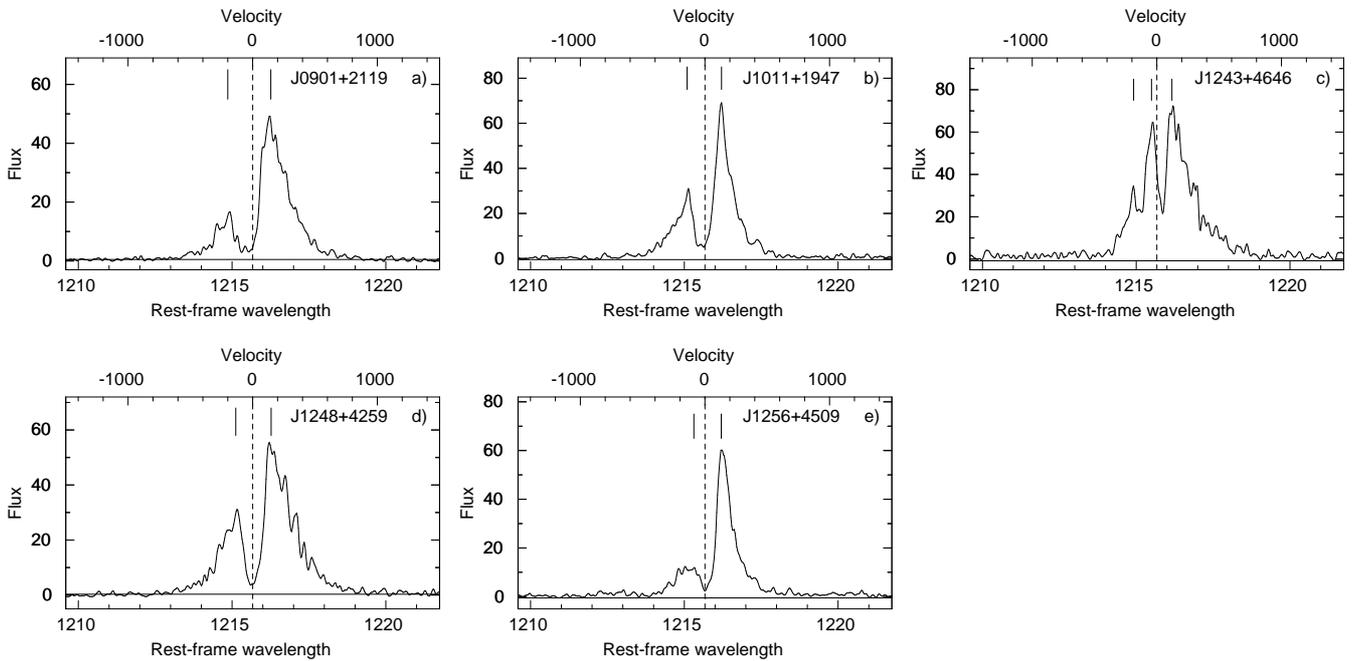

\hbox{
\includegraphics[angle=-90,width=0.33\linewidth]{spectrumLa_J0901+2119_c.ps}
\includegraphics[angle=-90,width=0.33\linewidth]{spectrumLa_J1011+1947_c.ps}
%
\includegraphics[angle=-90,width=0.33\linewidth]{spectrumLa_J1243+4646_c.ps}
}
\hbox{
\includegraphics[angle=-90,width=0.33\linewidth]{spectrumLa_J1248+4259_c.ps}
%
\includegraphics[angle=-90,width=0.33\linewidth]{spectrumLa_J1256+4509_c.ps}
}
\caption{Ly$\alpha$ profiles. Vertical dashed lines and short vertical solid
lines indicate the centres of profiles and profile peaks, respectively.
Flux densities are in 10$^{-16}$ erg s$^{-1}$ cm$^{-2}$\AA$^{-1}$, wavelengths are 
in \AA\ and velocities are in km s$^{-1}$. \label{fig7}}
\end{figure*}

\section{Ly$\alpha$ emission}\label{sec:lya}

One of the goals of our project is to search empirically for a possible 
correlation between the amount of escaping LyC radiation and the shape of the 
Ly$\alpha$ line profile. 
\citet{V17} have proposed that the presence of a double-peaked Ly$\alpha$ 
profile, with a small peak separation, would be a good indicator of 
LyC leakage. According to the models 
of \citet{V15}, the peak separation decreases with decreasing column 
density of the neutral gas. This in turn would result in a higher escape
fraction of the LyC radiation.

A strong Ly$\alpha$ $\lambda$1216\,\AA\ emission-line is detected in the 
medium-resolution spectra of all our galaxies (Fig. \ref{fig7}). 
In general, the profiles show two peaks, labelled by two short 
vertical lines. This shape is similar to that observed in known LyC 
leakers \citep{V17} and in some other galaxies \citep{JO14,H15,Y17}.
The Ly$\alpha$ profile of J1243$+$4646 is more
complex and consists of three peaks. Interestingly, it is quite similar to that
of the $z$ = 3.999 LyC leaker {\em Ion3} recently discovered by \citet{Va18}
and of the lensed galaxy Sunburst Arc at $z$ = 2.4 \citep{RT17}.
Some parameters of Ly$\alpha$ emission
are presented in Table \ref{tab7}. For J1243$+$4646, two separations are
given. It is seen that the separation between peaks varies in a range between
$\sim$ 150 km s$^{-1}$ and $\sim$ 350 km s$^{-1}$ with the lowest values
for J1243$+$4646 and J1256$+$4509. These are also among the lowest values
found for low-redshift LyC leakers.
The Ly$\alpha$ profiles of these two galaxies indicate that they are
efficient LyC leakers, similar to the galaxy J1154$+$2443 \citep{I18}.
This is examined quantitatively in Sect.~\ref{Ind}.

\begin{figure*}
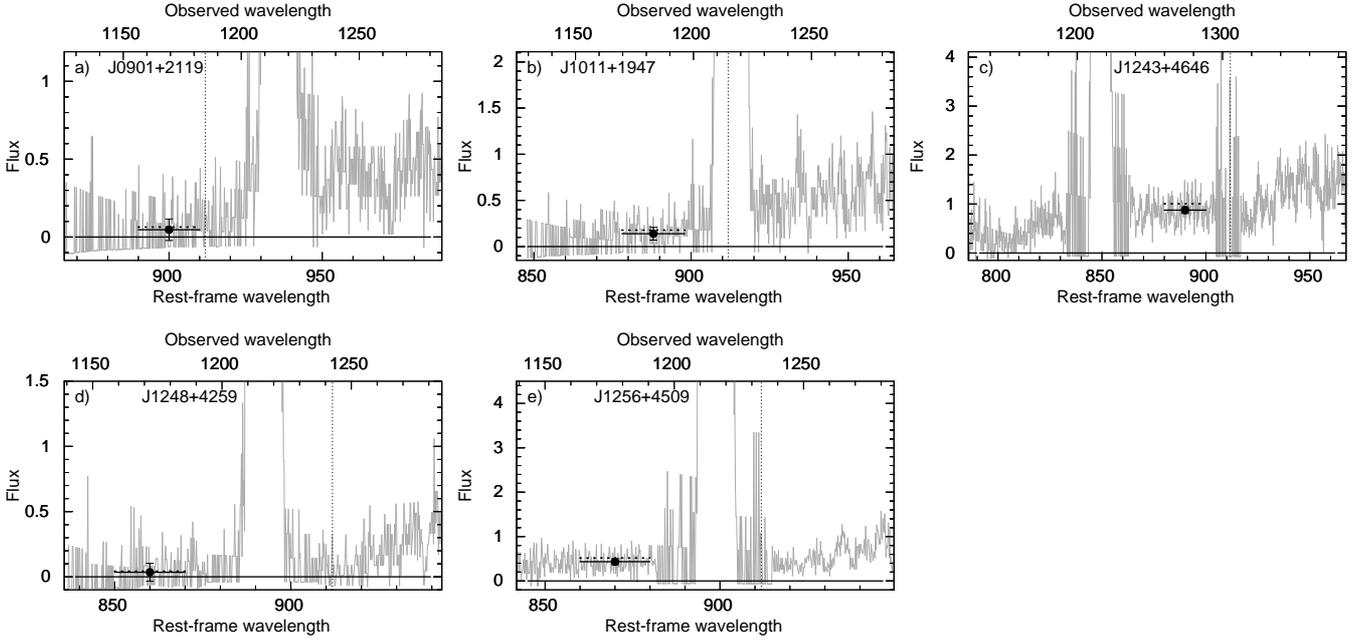

\hbox{
\includegraphics[angle=-90,width=0.33\linewidth]{spectrumLc_J0901+2119_expand_c.ps}
\includegraphics[angle=-90,width=0.33\linewidth]{spectrumLc_J1011+1947_expand_c.ps}
\includegraphics[angle=-90,width=0.33\linewidth]{spectrumLc_J1243+4646_expand_c.ps}
}
\hbox{
\includegraphics[angle=-90,width=0.33\linewidth]{spectrumLc_J1248+4259_expand_c.ps}
\includegraphics[angle=-90,width=0.33\linewidth]{spectrumLc_J1256+4509_expand_c.ps}
}
\caption{Segments of COS G140L spectra with the Lyman continuum.  Filled
circles show the average observed values with the 3$\sigma$ error bars.
The observed mean LyC flux densities and flux densities after
correction for the Milky Way extinction are indicated by short solid 
horizontal lines and short dotted horizontal lines, respectively. 
The Lyman limit at  912\AA\  rest-frame wavelength is indicated by dotted
vertical lines. Zero flux densities are shown by horizontal lines.
Flux densities are in 10$^{-16}$ erg s$^{-1}$ cm$^{-2}$\AA$^{-1}$, wavelengths 
are in \AA.
\label{fig8}}
\end{figure*}

  \begin{table*}
  \caption{LyC escape fraction \label{tab8}}
\begin{tabular}{lccccccrr} \hline
Name&$\lambda_0$$^{\rm a}$&$A$(LyC)$_{\rm MW}$$^{\rm b}$&$I_{\rm mod}$$^{\rm c,d}$&$I_{\rm obs}$(total)$^{\rm c,e}$&$I_{\rm obs}$(shadow)$^{\rm c,f}$&$I_{\rm esc}$(total)$^{\rm c,g}$&\multicolumn{1}{c}{$f_{\rm esc}$$^{\rm h}$}&\multicolumn{1}{c}{$f_{\rm esc}$$^{\rm i}$} \\
    &(\AA)&(mag)&&&&&\multicolumn{1}{c}{(per cent)}&\multicolumn{1}{c}{(per cent)} \\
\hline
J0901$+$2119&890-910&0.280&2.47$\pm$0.22&0.055$\pm$0.013&0.039$\pm$0.018&0.067$\pm$0.016& 2.7$\pm$0.7& 2.1$\pm$0.3 \\
J1011$+$1947&878-898&0.268&1.65$\pm$0.20&0.146$\pm$0.015&0.113$\pm$0.023&0.187$\pm$0.019&11.4$\pm$1.8& 6.2$\pm$0.7 \\
J1243$+$4646&880-900&0.118&1.29$\pm$0.17&0.874$\pm$0.020&0.914$\pm$0.096&0.936$\pm$0.021&72.6$\pm$9.7&72.3$\pm$7.2 \\
J1248$+$4259&850-870&0.235&2.30$\pm$0.15&0.040$\pm$0.012&0.033$\pm$0.015&0.050$\pm$0.015& 2.2$\pm$0.7& 1.4$\pm$0.4 \\
J1256$+$4509&860-880&0.197&1.38$\pm$0.20&0.438$\pm$0.019&0.326$\pm$0.068&0.525$\pm$0.022&38.0$\pm$5.7&46.0$\pm$5.6 \\ 
\hline
  \end{tabular}

\hbox{$^{\rm a}$Rest-frame wavelength range in \AA\ used for the determination of the LyC flux density.}

\hbox{$^{\rm b}$Milky Way extinction at the mean observed wavelengths of the 
range used for the determination of the LyC flux density.} 

\hbox{\, The \citet{C89} reddening law with $R(V)$ = 3.1 is adopted.} 

\hbox{$^{\rm c}$in 10$^{-16}$ erg s$^{-1}$cm$^{-2}$\AA$^{-1}$.}

\hbox{$^{\rm d}$LyC flux density derived from the modelled SED.}

\hbox{$^{\rm e}$Observed LyC flux density derived from the spectrum with total exposure.}

\hbox{$^{\rm f}$Observed LyC flux density derived from the spectrum with shadow exposure.}

\hbox{$^{\rm g}$Observed LyC flux density in the spectrum with total exposure which is corrected for the Milky Way 
extinction.}

\hbox{$^{\rm h}$$I_{\rm esc}$(total)/$I_{\rm mod}$, where $I_{\rm mod}$ is derived from SED (first method).}

\hbox{$^{\rm i}$$I_{\rm esc}$(total)/$I_{\rm mod}$, where $I_{\rm mod}$ is derived from H$\beta$ flux density (second method).}

  \end{table*}

\begin{figure*}
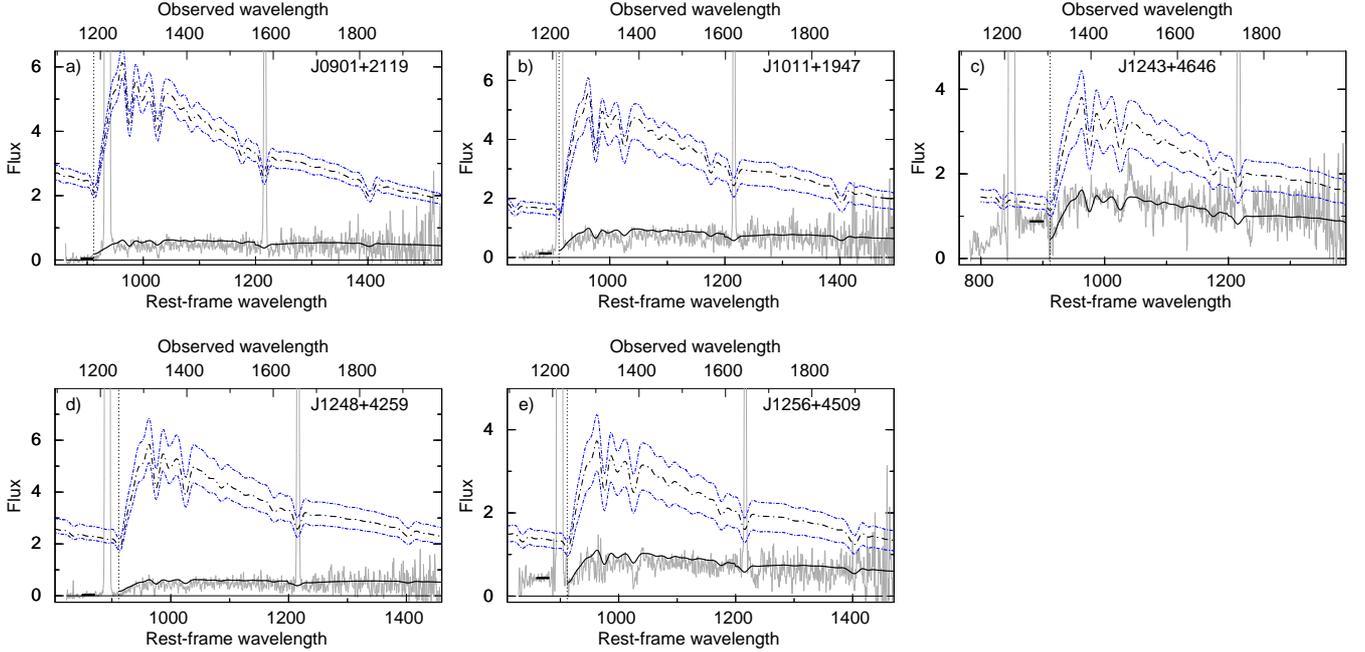

\hbox{
\includegraphics[angle=-90,width=0.33\linewidth]{spectrumLc_J0901+2119_c.ps}
\includegraphics[angle=-90,width=0.33\linewidth]{spectrumLc_J1011+1947_c.ps}
\includegraphics[angle=-90,width=0.33\linewidth]{spectrumLc_J1243+4646_c.ps}
}
\hbox{
\includegraphics[angle=-90,width=0.33\linewidth]{spectrumLc_J1248+4259_c.ps}
%
\includegraphics[angle=-90,width=0.33\linewidth]{spectrumLc_J1256+4509_c.ps}
}
\caption{COS G140L spectra of our sources (grey lines)
superposed by the modelled SEDs (thick solid lines), reddened by 
both the internal and Milky Way extinctions.  
The unreddened (intrinsic) SEDs are shown by the thick dash-dotted lines.
The variations of intrinsic SEDs are shown by thin dash-dotted lines for 
1$\sigma$ uncertainties of $C$(H$\beta$)$_{\rm int}$
(see Table \ref{tab3}).
$R(V)$$_{\rm int}$ = 2.4 and $R(V)$$_{\rm MW}$ = 3.1 are adopted in all panels
except for the galaxy J1243$+$4646 for which $R(V)$$_{\rm int}$ = 3.1 is 
adopted. 
Lyman limit at the rest-frame wavelength 912\AA\ is indicated by dotted
vertical lines. Zero flux densities are represented by solid horizontal lines.
Flux densities are in 10$^{-16}$ erg s$^{-1}$ cm$^{-2}$\AA$^{-1}$, wavelengths 
are in \AA. \label{fig9}}
\end{figure*}

\section{Escaping Lyman continuum radiation}\label{sec:lyc}

\subsection{Observed and intrinsic Lyman continuum flux densities}

Blow-ups of the LyC spectral region are illustrated in Fig.~\ref{fig8} showing clearly 
that  Lyman continuum emission is significantly detected in the spectra of all galaxies.
The average flux densities are shown by filled circles with 3$\sigma$ statistical error bars. 
The short thick solid lines indicate wavelength ranges used for averaging.
The observed LyC emission flux densities $I_{\rm obs}$(total) should  be corrected for the 
Milky Way extinction to derive the escaping flux densities $I_{\rm esc}$(total).
The extinction-corrected average LyC flux densities $I_{\rm esc}$(total) are shown by the 
dotted horizontal lines in Fig. \ref{fig8}. The corresponding measurements are summarized
in Table \ref{tab8}.

We also
show in Table \ref{tab8} the observed LyC flux densities $I_{\rm obs}$(shadow)
derived from the exposures during periods with considerably lower 
night sky emission. In general, $I_{\rm obs}$(shadow) is somewhat lower than
$I_{\rm obs}$(total), implying some contribution of the geocoronal scattered
light to the flux densities in observations with the total exposures.
In any case both measurements are consistent within the 
2$\sigma$ errors for all galaxies. 

\begin{figure*}
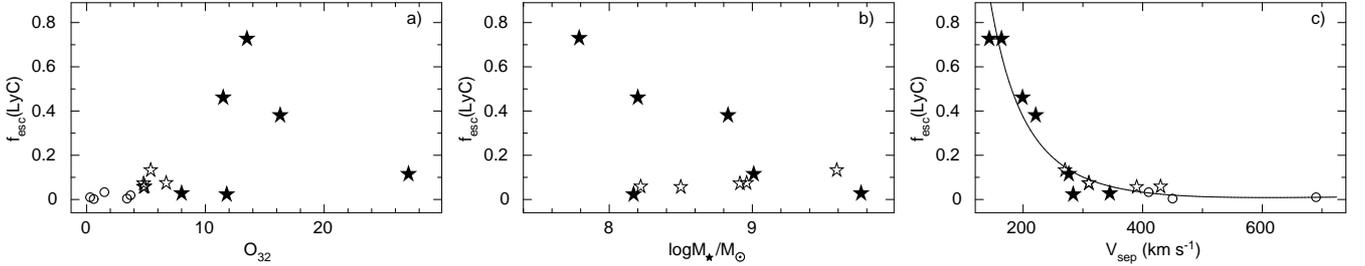

\hbox{
\includegraphics[angle=-90,width=0.33\linewidth]{lyco32_2.ps}
\includegraphics[angle=-90,width=0.33\linewidth]{lycmstar.ps}
\includegraphics[angle=-90,width=0.33\linewidth]{lycsep_2.ps}
}
\caption{Relations between the Lyman continuum escape fraction
$f_{\rm esc}$(LyC) in low-redshift LyC leaking galaxies and {\bf a)} the 
[O~{\sc iii}]$\lambda$5007/[O~{\sc ii}]$\lambda$3727 
emission-line ratio, {\bf b)} the stellar mass $M_\star$, and
{\bf c)} the separation $V_{\rm sep}$ between the Ly$\alpha$ profile peaks. 
LyC leakers from \citet{I18} and this paper are shown by filled stars, 
the LyC leakers from \citet{I16,I16b} are represented by open stars
and other LyC leaking galaxies from \citet{L13}, \citet{B14} and \citet{L16} 
with $f_{\rm esc}$(LyC) derived by \citet{C17} are shown by open circles.
The two separations in the galaxy J1243$+$4646 with the highest 
$f_{\rm esc}$(LyC) are shown (see Fig. \ref{fig7}c and Table \ref{tab7}).
\label{fig10}}
\end{figure*}


To derive the escape fraction of ionizing radiation in our sources we use two
methods to determine the intrinsic emission in the Lyman continuum, 
following \citet{I16,I16b,I18}.  We here briefly illustrate the steps  of the first method, which uses the SED fits combining the 
SDSS optical spectra and the COS UV spectra described above to predict the LyC radiation.

The observed G140L total-exposure spectra (grey lines) and predicted intrinsic SEDs  (thick dash-dotted lines) 
are shown in Fig. \ref{fig9}, together with the  reddened intrinsic SEDs (thick solid lines).
The predicted SEDs are
obtained from fitting optical SDSS spectra which are corrected for
the Milky Way extinction at observed wavelengths, adopting  $A(V)$(MW) from 
the NED and for the internal extinction at rest-frame wavelengths, 
the preferred UV attenuation law (i.e.\ $R_V$ discussed above), and the
extinction coefficients $C$(H$\beta$)$_{\rm int}$ derived from the 
hydrogen Balmer decrement.


The thin dash-dotted lines show
the intrinsic SEDs derived by adopting extinction coefficients 
$C$(H$\beta$)$_{\rm int}$ $\pm$ $\sigma$[$C$(H$\beta$)$_{\rm int}$] where 
$\sigma$[$C$(H$\beta$)$_{\rm int}$]
are 1$\sigma$ errors of $C$(H$\beta$)$_{\rm int}$. It is seen that the intrinsic 
LyC flux density depends on the adopted extinction coefficient, being higher 
for higher $C$(H$\beta$)$_{\rm int}$. 
This is due to the fact that a higher 
extinction-corrected
H$\beta$ flux density corresponds to a higher number of ionizing photons.

\begin{figure*}
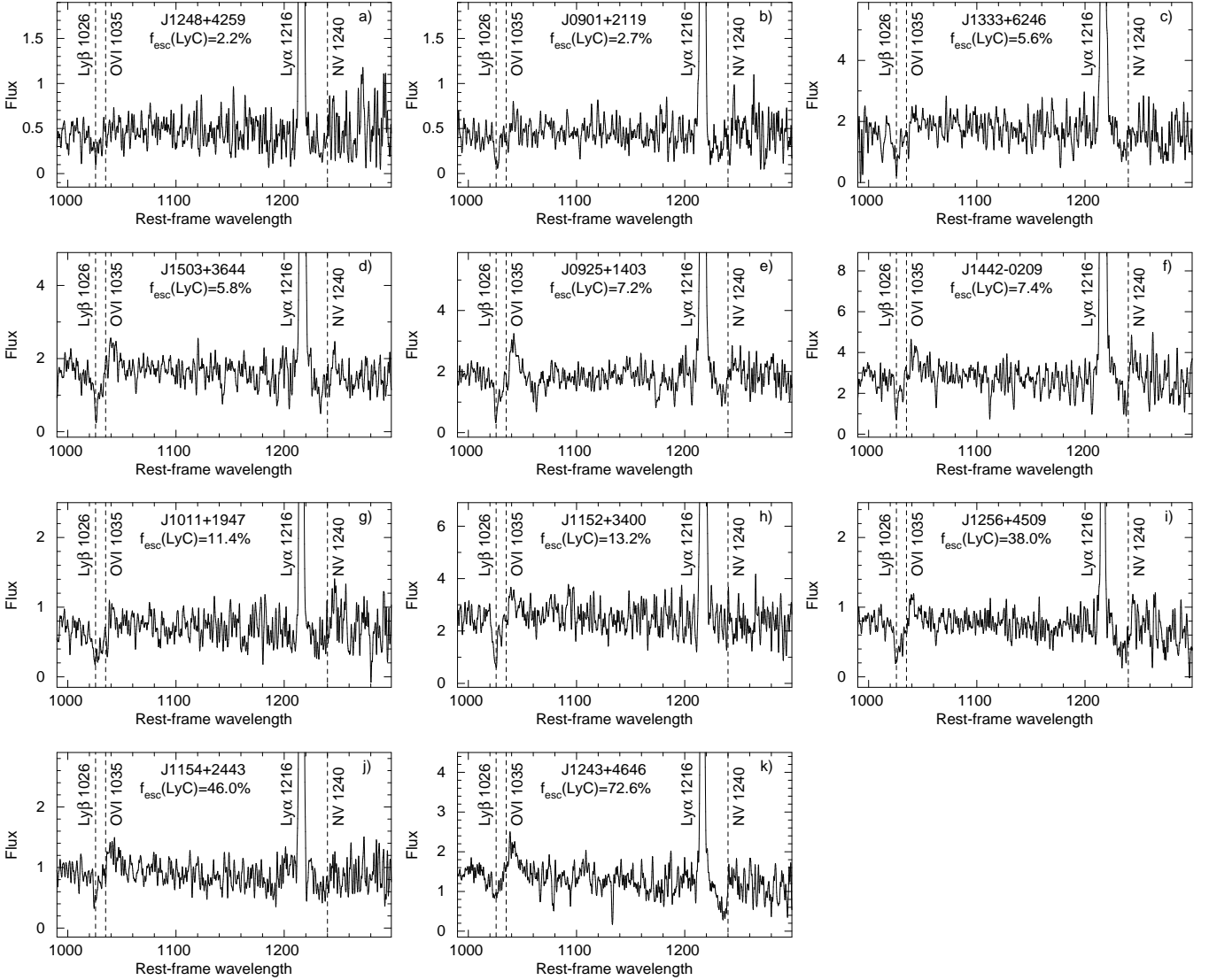

\hbox{
\includegraphics[angle=-90,width=0.33\linewidth]{stellar_J1248+4259_c.ps}
\includegraphics[angle=-90,width=0.33\linewidth]{stellar_J0901+2119_c.ps}
\includegraphics[angle=-90,width=0.33\linewidth]{stellar_J1333+6246_c.ps}
}
\hbox{
\includegraphics[angle=-90,width=0.33\linewidth]{stellar_J1503+3644_c.ps}
\includegraphics[angle=-90,width=0.33\linewidth]{stellar_J0925+1403_c.ps}
\includegraphics[angle=-90,width=0.33\linewidth]{stellar_J1442-0209_c.ps}
}
\hbox{
\includegraphics[angle=-90,width=0.33\linewidth]{stellar_J1011+1947_c.ps}
\includegraphics[angle=-90,width=0.33\linewidth]{stellar_J1152+3400_c.ps}
\includegraphics[angle=-90,width=0.33\linewidth]{stellar_J1256+4509_c.ps}
}
\hbox{
\includegraphics[angle=-90,width=0.33\linewidth]{stellar_J1154+2443_c.ps}
\includegraphics[angle=-90,width=0.33\linewidth]{stellar_J1243+4646_c.ps}
}
\caption{Segments of COS G140L spectra with broad stellar lines 
O {\sc vi} $\lambda$1035\AA\ and N {\sc v} $\lambda$1240\AA\ for all
LyC leakers studied by \citet{I16,I16b,I18} and in this paper. The centres of
these lines and of the Ly$\beta$ line are indicated by vertical dashed lines.
The spectra are ordered according to the $f_{\rm esc}$(LyC) values from lowest 
(top) to highest (bottom).
Flux densities are in 10$^{-16}$ erg s$^{-1}$ cm$^{-2}$\AA$^{-1}$, wavelengths 
are in \AA.
\label{fig11}}
\end{figure*}

\subsection{LyC escape fraction}

To derive $f_{\rm esc}$(LyC)
\citet{I16,I16b} and \citet{I18} used the ratio of the escaping
flux densities $I_{\rm esc}$ to the intrinsic flux densities $I_{\rm mod}$
of the Lyman continuum:
\begin{equation}
f_{\rm esc}({\rm LyC}) =\frac{I_{\rm esc}(\lambda)}{I_{\rm mod}(\lambda)}, 
\label{eq:fesc}
\end{equation}
where $\lambda$ is the mean wavelength of the range used for averaging of
the LyC flux density (see Table \ref{tab8}).
%
They proposed two methods to derive the intrinsic flux densities $I_{\rm mod}$.
The first method is based on the SED fitting as described above.
The relation between the luminosities of hydrogen recombination lines 
and the number of ionizing photons emitted per unit time, $N$(Lyc),
is used in the second method. 
The details of both methods are described by \citet{I18}.

Using Eq. \ref{eq:fesc}, we derive the escape fractions 
$f_{\rm esc}$(LyC) by both methods accounting for the uncertainties of the
observed monochromatic LyC flux densities and those of the extinction
coefficients $C$(H$\beta$)$_{\rm int}$.
For most of our sources the two methods give escape fractions which agree
within the uncertainties.
The derived $f_{\rm esc}$(LyC) (Table \ref{tab8}) vary over a wide
range and are very high
for J1243$+$4646 and J1256$+$4509 with escape fractions of $\sim 72$ per cent and 38 per cent, respectively. 
These values, together with $f_{\rm esc}$(LyC) = 46 per cent
for J1154$+$2443 \citep{I18}, are among the highest known for low- and
high-redshift galaxies. They are far above the average LyC escape fractions 
for SFGs of $\sim$ 10 -- 20 per cent needed to 
fully reionize the Universe at redshifts $z$ $>$ 5.


\subsection{Indicators of high $f_{\rm esc}$(LyC)} \label{Ind}

The direct detection of LyC emission in low-redshift star-forming galaxies
is a difficult task. At the moment, only {\sl HST} can be used for that 
purpose. Although our targets are relatively bright in the UV ($M_{\rm UV}$ 
close to $M_{\rm UV}^*$ of the
high-$z$ luminosity function) and the COS G140L observations are fairly short
(1 -- 2 orbits), relatively few galaxies at $z$~$\ga$~0.3 can be observed 
directly in the LyC in this manner. 
Therefore, reasonable indirect indicators of LyC leakage are needed to build a 
larger sample for statistical studies. We consider below some possible 
indicators.

\citet{G04}, \citet{JO13} and \citet{NO14} proposed that a high
O$_{32}$ ratio can be an indication of density-bounded H~{\sc ii} regions and
thus of high $f_{\rm esc}$(LyC). At low redshifts the [O~{\sc ii}]$\lambda$3727 
and [O~{\sc iii}]$\lambda$5007 emission lines are seen in the optical range.
Their flux densities are available for large samples of SFGs. 
Now with our new LyC leakers we have in hand
a sample of sixteen galaxies with a wide range of O$_{32}$ $\sim$ 0.5 -- 27
\citep[][ this paper]{L13,B14,L16,I16,I16b,I18}. The relation between 
$f_{\rm esc}$(LyC) and O$_{32}$ is presented in Fig. \ref{fig10}a.
Although there is a trend of increasing $f_{\rm esc}$(LyC) with increasing 
of O$_{32}$, the spread of $f_{\rm esc}$(LyC) at high O$_{32}$ is large. 
Furthermore, the LyC
escape fraction in the galaxy J1011$+$1947 with the highest O$_{32}$ is 
relatively low. This is in line with the conclusions of \citet{I17} that
a high O$_{32}$ is a necessary but not sufficient condition for escaping 
ionizing radiation.  The O$_{32}$ ratio
also depends on other parameters, such as ionization parameter, 
hardness of ionizing radiation and metallicity of the H~{\sc ii} region.
The spread of galaxies in the $f_{\rm esc}$(LyC) -- O$_{32}$ diagram can also be
caused by their different orientations relative to the observer and by
inhomogeneous LyC leakage. For example, it was demonstrated by \citet{TI97} that
the low-metallicity galaxy SBS 0335$-$052E with a high O$_{32}$ $\sim$ 15
shows very strong Ly$\alpha$ absorption in its UV spectrum, indicating
an extremely high neutral hydrogen column density $N$(H~{\sc i}) 
$\sim$ 7$\times$10$^{21}$ cm$^{-2}$ along the line of sight, preventing 
leakage of LyC radiation. However, recently \cite{H17} discovered
holes in the H~{\sc i} envelope of this galaxy probably oriented nearly 
perpendicularly
to the line of sight, which could allow for LyC leakage in some directions.

It has also been suggested that $f_{\rm esc}$(LyC) tends to be higher in 
low-mass galaxies \citep{W14,T17}. Stellar masses $M_\star$ 
are available for a large number of star-forming galaxies because observations 
with ground-based telescopes are sufficient for their determination. In 
particular, SDSS spectra can be used for that. We show in Fig.~\ref{fig10}b
the relation between $f_{\rm esc}$(LyC) and stellar mass $M_\star$. It is seen
that there is a tendency of increasing $f_{\rm esc}$(LyC) with decreasing 
$M_\star$, although a large scatter is present.
New observations of LyC 
leakers with lower log $M_\star$/$M_\odot$ $<$ 8.0 are needed to verify the
suggested increase of $f_{\rm esc}$(LyC) with decreasing $M_\star$.

The profile of the Ly$\alpha$ emission line can also be used as an indirect
indicator of the LyC leakage. This indicator is most useful because it can be 
applied to nearby low-mass galaxies for which direct observations of the LyC 
are not possible because of their low redshift. 
\citet{V17} and \citet{I18} found a tight dependence of $f_{\rm esc}$(LyC)
on the separation $V_{\rm sep}$ between the peaks of the Ly$\alpha$
emission line in LyC leakers. Adding the new LyC leakers discussed in this 
paper extends the relation to higher $f_{\rm esc}$(LyC) and lower $V_{\rm sep}$ 
and strengthens it considerably (Fig. \ref{fig10}c). The regression line 
(solid line) to this relation is
\begin{equation}
f_{\rm esc}({\rm LyC}) = \frac{3.23\times 10^4}{V{_{\rm sep}}^2} - 
\frac{1.05\times 10^2}{V{_{\rm sep}}} + 0.095, \label{eq:fescVsep}
\end{equation}
where $V_{\rm sep}$ is in km s$^{-1}$. We note that the relation  
Eq. \ref{eq:fescVsep} is empirical and is incorrect at very
small Ly$\alpha$ peak separations resulting in $f_{\rm esc}$(LyC) $>$ 1 at
$V_{\rm sep}$ $\la$ 140 km s$^{-1}$.

Both the LyC and Ly$\alpha$ escaping radiation are determined by the
column density of the neutral gas in LyC leaking galaxies
\citep[e.g. ][]{V15,V17} implying a tight relation between $V_{\rm sep}$
and $f_{\rm esc}$(LyC), as it is in Fig. \ref{fig10}c. 
This relation described by Eq. \ref{eq:fescVsep} should constitute a strong 
constraint for constructing  radiative transfer and kinematical models which 
simultaneously reproduce $f_{\rm esc}$(LyC) and the Ly$\alpha$ profile.

Another potential indicator is shown in Fig. \ref{fig11}, where we
present segments of COS
G140L spectra of the LyC leakers we have studied previously and in this
paper,
in the wavelength range $\sim$ 1000 -- 1300\AA\
which includes the stellar lines O~{\sc vi} $\lambda$1035 and
N~{\sc v} $\lambda$1240 with P-Cygni profiles, indicative of stellar
winds from hot
massive stars \citep[this paper, ][]{I16,I16b,I18}. The spectra are
shown in
order of increasing $f_{\rm esc}$(LyC). The stellar lines are very
weak in spectra of galaxies with lowest $f_{\rm esc}$(LyC) $\la$ 5 per cent
(Figs. \ref{fig11}a-c). In spectra of all other LyC leakers with higher
$f_{\rm esc}$(LyC) these lines are clearly seen, implying that,
in principle, stellar lines with P-Cygni profiles from hot massive stars
could
also be considered as indicators of LyC leakage.

Finally, a  coherent analysis of UV absorption lines (including H lines of the 
Lyman series and metallic lines such as Si~{\sc ii}\,$\lambda1260$) and the UV
attenuation also provide a consistent and accurate measure of the Lyman 
continuum escape fraction, as shown in detail by \citet{G18} and \cite{C18}.

Overall we conclude that the most reliable and simple empirical indicator
of LyC leakage is the separation between the peaks of the Ly$\alpha$
profile. There are, however, several problems with the use of this
indicator. First, the Ly$\alpha$
line must be resolved for the study of its profile. Additionally, for
low-redshift galaxies, UV observations are needed. Finally, at redshifts
corresponding to the epoch of recombination, Ly$\alpha$ is effectively
absorbed by the intergalactic neutral medium.

\section{Discussion}\label{discussion}

We now briefly compare our sources with those studied in other recent papers.
\citet{He18} have observed seven $z \sim 0.3$ star-forming galaxies with
the COS spectrograph targeting also the Lyman continuum. No LyC emission is detected,
translating to an absolute LyC escape fraction of $f_{\rm esc}({\rm LyC}) =0.4^{+10}_{-0.4}$ per cent
from the stacked (combined) spectrum, excluding thus strong LyC leakage.
The galaxies were selected from the COSMOS survey as star-forming galaxies, applying a few additional 
selection criteria. Both the rest-frame UV (\lya) and optical emission lines of their sources are 
relatively weak compared to our galaxies, 
as measured by their equivalent widths which are quite low, EW(H$\beta$) $\sim 7-77$ \AA,
compared to our median value of $\sim 220$ \AA. 
Correspondingly, the specific star formation rate sSFR of these galaxies is fairly modest
(sSFR $ \sim 10^{-9.7}$ to $10^{-8.1}$ yr$^{-1}$). 
Comparing these properties with those of our sources, the LyC non-detection in 
the galaxies of \citet{He18} may not be surprising. Indeed, the strong LyC 
leakers detected so far are all characterized by their compactness, very strong 
emission lines and very high sSFR. These properties are also typical of those of
normal $z>6$ star-forming galaxies, making our compact leakers good analogs
of the sources of cosmic reionization, as shown by \citet[cf.][]{S16}, whereas 
the galaxies observed by \citet{He18} are more typical of low-$z$ 
non-compact star-forming galaxies.

A recent study by \cite{N18} has examined Lyman continuum escape in 
$z \sim 3.5$ galaxies with strong [O~{\sc iii}] emission lines, finding a low 
escape fraction of $<10$ per cent and questioning the usefulness of high 
O$_{32}$ ratios for finding strong LyC emitters. Using an overdensity at 
$z =3.42-3.58$, these authors have selected 73 star-forming galaxies 
to search for LyC emission probed by a deep VIMOS $U$-band image. Again, the 
LyC remains undetected both for the individual sources and in stacks, from which
they derive a $1 \sigma$ upper limit of $\fesc$(LyC) $< 6.3 \pm 0.7$ per cent
on the relative LyC escape fraction of the total sample.
They also examine  ``strong  [O~{\sc iii}]'' emitters, finding 
$\fesc$(LyC) $< 8.2 \pm 0.8$ per cent for this subsample of 54 sources with rest-frame 
EW(\oiiil) $\sim 400$ \AA\ and an estimated 
O$_{32} \sim 4.3$. From this, \cite{N18} conclude that their result raises 
questions about the reliability of extreme EW(\oiiil)  and O$_{32}$ as 
effective tracers of LyC escape.

A quick comparison with the properties of the $z \sim 0.3$ galaxies studied in our work
\citep[][and the present work]{I16,I18} shows that our sources have significantly stronger
emission lines (EW(\oiiill ) $\sim 1400-2100$ \AA\ compared to $\sim$ 400 \AA\ rest-frame) and 
higher O$_{32}$ line ratios (extinction-corrected O$_{32} \ga 5$ compared to $\sim 4$ 
before extinction correction). At O$_{32} < 4$ we do not find any source with 
absolute LyC escape fractions above $\sim 2$ per cent, which is compatible with 
their non-detections. 
In short, the result of  \cite{N18} does not contradict our findings and hence does not exclude 
that galaxies with very strong emission lines and a high ratio O$_{32}$ are Lyman continuum emitters,
as we found at $z \sim 0.3$, although there is no tight correlation between 
O$_{32}$ and LyC escape in our sample.

Another recent study examined the LyC escape fraction of emission line-selected $z \sim 2.5$ galaxies,
finding  1 $\sigma$ upper limits on the absolute escape fraction of $\fesc < 6 - 14$ per cent \citep{R17},
where the latter value is applicable to a subsample of 13 sources with O$_{32} >5$.
As mentioned by these authors, their observations are not deep enough to detect 
escape fractions of $\sim 2 - 15$ per cent as measured for the majority of our 
sources (8 out of 11).
On the other hand, the three galaxies with  the highest LyC escape fractions in our sample have
all fairly low stellar masses $M_\star \la 10^9$ \msun, clearly outside the mass range of the galaxies
observed by \citet{R17}, which have masses $M_\star \sim 10^9 - 10^{10.5}$ \msun, 
with a median mass of $M_\star \sim 10^{9.7}$ \msun.
Hence their results are also compatible with our findings.
For comparison, the stellar masses of the low-$z$ galaxies observed by \citet{He18} 
span a fairly large range 
with the bulk of the sources between $M_\star \sim 10^{9.5} - 10^{10}$ \msun, as 
found from the catalog of \citet{La16}.
These results combined with our findings indicate that Lyman continuum leakage is probably
more common and higher in galaxies of low mass.

From the available observations, we conclude that compact and intensely star-forming galaxies with
sufficiently strong optical emission lines and a high O$_{32}$ line ratio are excellent 
Lyman continuum leaker candidates.
Furthermore, the $z \sim 0.3$ galaxies we have studied are outliers of the galaxy population at low redshift, 
but their properties are fairly typical of galaxies in the early Universe \citep[cf.][]{S16}.
It is therefore straightforward to suggest that the average Lyman continuum escape fraction 
of a galaxy at a given stellar mass should increase with increasing redshift, as ultimately needed
to explain cosmic reionization. In addition, LyC escape could also be more frequent and/or
the average escape fraction higher in low mass galaxies 
(10$^7$ $<$ $M_\star < 10^9$ \msun),
as suggested by numerical simulations \citep[e.g.][]{W14,T17}.

\section{Conclusions}\label{summary}

We present new {\sl Hubble Space Telescope} ({\sl HST}) Cosmic
Origins Spectrograph (COS) observations of five 
compact star-forming galaxies (SFG) with high O$_{32}$ = 
[O~{\sc iii}]$\lambda$5007/[O~{\sc ii}]$\lambda$3727 flux density ratios in 
the range $\sim$ 8 -- 27, and in the redshift range $z$ = 0.2993 -- 0.4317.
We use these data to study the Ly$\alpha$ emission and the escaping Lyman 
continuum (LyC) radiation of these SFGs. This study is an extension of the work
reported earlier in \citet{I16,I16b,I18}.
Our main results are summarized as follows:

1. LyC leakage, i.e.\ emission of Lyman continuum radiation, is detected in all five galaxies. The highest
escape fractions, $f_{\rm esc}$(LyC) = 72~$\pm$~10 per cent and 38~$\pm$~6 per 
cent, are found in J1243$+$4646 and J1256$+$4509, respectively. These values are
among the highest known for LyC leakers at any redshift.

2. A Ly$\alpha$ emission line with two peaks is observed in the
spectra of four galaxies, and a Ly$\alpha$ emission line with three peaks is 
detected in the spectrum of J1243$+$4646, the object with the highest 
$f_{\rm esc}$(LyC). A strong anti-correlation between $f_{\rm esc}$(LyC) and the 
peak velocity separation of the Ly$\alpha$ profile is found, as suggested earlier
by \citet{I18}.

3. We find that a high O$_{32}$ ratio is a necessary but not sufficient
condition for a large amount of Lyman continuum radiation escaping from SFGs. 
Instead, the most reliable selection criterion for LyC leakers is the 
peak separation of the Ly$\alpha$ profile $V_{\rm sep}$ which is strongly 
anti-correlated to the LyC escape fraction $f_{\rm esc}$(LyC),
as qualitatively predicted by the radiation transfer simulations  of \citet{V15}.

4. A bright compact star-forming region 
(with the exception of J0901+2119 which shows 2 fainter knots of star formation 
in addition to the main star-forming region) is seen
in the COS near ultraviolet (NUV) acquisition images. The surface brightness
at the outskirts can be approximated by an exponential disc, with a scale 
length of $\sim$ 1.48 -- 1.75 kpc, indicating 
that all our LyC leakers may be dwarf disc systems. 

5. Our galaxies are characterized by stellar masses $M_\star$ 
$\sim$ 10$^{7.8}$ -- 10$^{9.8}$~M$_\odot$ and high star-formation rates in the 
range SFR $\sim$ 18 -- 80~M$_\odot$~yr$^{-1}$. Their specific star formation rates
are among the highest found in low-redshift LyC leakers. The metallicities of 
our galaxies are low, ranging from 12 + logO/H = 7.64 to 8.16.


We have discovered a class of galaxies in the local Universe which are
leaking ionizing radiation, some with very high escape fractions ($\ga$
40 per cent). They share many properties of high-redshift galaxies, indicating
that stars in high-redshift galaxies may have reionized the Universe.
The dominance of very young ($\la 3$ Myr) stars in their intense
star-forming events suggests that stellar winds and radiation from massive
stars may be an important source driving the escape of ionizing radiation.
These local galaxies are ideal laboratories for pursuing the
investigation of the mechanisms responsible for the
escape of ionizing radiation from galaxies.

\section*{Acknowledgements}

Based on observations made with the NASA/ESA {\sl Hubble Space Telescope}, 
obtained from the data archive at the Space Telescope Science Institute. 
STScI is operated by the Association of Universities for Research in Astronomy,
Inc. under NASA contract NAS 5-26555. Support for this work was provided by 
NASA through grant number HST-GO-14635.002-A from the Space Telescope Science 
Institute, which is operated by AURA, Inc., under NASA contract NAS 5-26555.
G.W. has been supported by the Deutsches Zentrum f\"ur Luft- und Raumfahrt 
(DLR) through grant number 50OR1720.
I.O. acknowledges grants GACR 14-20666P and 17-06217Y of the Czech National Foundation. 
Funding for SDSS-III has been provided by the Alfred P. Sloan Foundation, 
the Participating Institutions, the National Science Foundation, and the U.S. 
Department of Energy Office of Science. The SDSS-III web site is 
http://www.sdss3.org/. SDSS-III is managed by the Astrophysical Research 
Consortium for the Participating Institutions of the SDSS-III Collaboration. 
GALEX is a NASA mission  managed  by  the  Jet  Propulsion  Laboratory.
This research has made use of the NASA/IPAC Extragalactic Database (NED) which 
is operated by the Jet  Propulsion  Laboratory,  California  Institute  of  
Technology,  under  contract with the National Aeronautics and Space 
Administration.





\begin{thebibliography}{}

\bibitem[Ade et al.(2014)]{P14} Ade P. A. R. et al.,
2014, \aap, 571, A16


\bibitem[Alam et al.(2015)]{A15}Alam S. et al., 2015, \apjs, 219, 12

\bibitem[Aller(1984)]{A84} Aller L. H., 1984, Physics of Thermal
Gaseous Nebulae. Dordrecht: Reidel


\bibitem[Baldwin et al.(1981)Baldwin, Phillips \& Terlevich]{BPT81} 
Baldwin J. A., Phillips M. M., Terlevich R., 1981, \pasp, 93, 5

\bibitem[Bian et al.(2017)]{B17} Bian F., Fan X., McGreer I., Cai Z., Jiang L.,
2017, \apj, 837, 12

\bibitem[Borthakur et al.(2014)]{B14} Borthakur S., Heckman T. M., 
Leitherer C., Overzier R. A., 2014, \sci, 346, 216



\bibitem[Bouwens et al.(2015)]{B15a} Bouwens R. J., Illingworth G. D., 
Oesch P. A., Caruana J., Holwerda B., Smit R.,  Wilkins S., 
2015, \apj, 811, 140

\bibitem[Bouwens et al.(2017)]{Bo17}  Bouwens R. J., Illingworth G. D., 
Oesch P. A., Atek H, Lam D, Stefanon M., 2017, \apj, 843, 41






\bibitem[Cardelli et al.(1989)Cardelli, Clayton \& Mathis]{C89} 
Cardelli J. A., Clayton G. C., Mathis J. S., 1989, \apj, 345, 245

\bibitem[Chisholm et al.(2017)]{C17} Chisholm J., Orlitov\'a I., 
Schaerer D., Verhamme A., Worseck G., Izotov Y. I., Thuan T. X., Guseva N. G.,
2017, \aap, 605, A67

\bibitem[Chisholm et al.(2018)]{C18} Chisholm J. et al.,
2018, \aap, in press; preprint arXiv:1803.03655


\bibitem[Curtis-Lake et al.(2016)]{CL16} Curtis-Lake E. et al., 
2016, \mnras, 457, 440

\bibitem[de Barros et al.(2016)]{B16} de Barros S. et al., 2016, \aap, 585, A51


\bibitem[Dressler et al.(2015)]{D15} Dressler A., Henry A., Martin C. L., 
Sawicki M., McCarthy P., Villaneuva E., 2015, \apj, 806, 19










\bibitem[Fisher et al.(2018)]{F18} Fischer W. J. et al., 2018, Cosmic Origins 
Spectrograph Instrument Handbook, Version 10.0. Baltimore: STScI


\bibitem[Gazagnes et al.(2018)]{G18} Gazagnes S., Chisholm J., Schaerer D., 
Verhamme A., Rigby J. R., Bayliss M., 2018, \aap, in press; preprint
arXiv:1802.06378






\bibitem[Grazian et al.(2016)]{Gr15} Grazian A. et al., 2016, \aap, 585, A48 

\bibitem[Green et al.(2012)]{Gr12} Green J. C. et al., 2012, \apj, 744, 60

\bibitem[Grimes et al.(2009)]{Gr09} Grimes J. P. et al., 2009, \apj, 181, 272

\bibitem[Guseva et al.(2004)]{G04} Guseva N. G., Papaderos P., Izotov Y. I., 
Noeske K. G., Fricke K. J., 2004, \aap, 421, 519




\bibitem[Guseva et al.(2013)]{G13} Guseva N. G., Izotov Y. I., Fricke K. J.,
Henkel C., 2013, \aap, 555, A90




\bibitem[Hassan et al.(2018)]{H18} Hassan S., Dav\'e R., Mitra S., 
Finlator K., Ciardi B., Santos M. G., 2018, \mnras, 473, 227





\bibitem[Henry et al.(2015)]{H15} Henry A., Scarlata C., Martin C. S., Erb D.,
2015, \apj, 809, 19

\bibitem[Herenz et al.(2017)]{H17} Herenz E. C., Hayes M., Papaderos P., 
Cannon J. M., Bik A. Melinder J., \"Ostlin G., 2017, \aap, 604, 99

\bibitem[Hernandez et al.(2018)]{He18} Hernandez S., Leitherer C., 
Boquien M., Buat V., Burgarella D., Calzetti D., Noll S., 2018,
\mnras, in press; preprint arXiv:1804.09721




\bibitem[Inoue et al.(2014)]{Inoue14} Inoue A.~K., Shimizu 
I., Iwata I., Tanaka M., 2014, \mnras, 442, 1805 





\bibitem[Izotov et al.(1994)Izotov, Thuan \& Lipovetsky]{ITL94} Izotov Y. I.,
Thuan T. X., Lipovetsky V. A., 1994, \apj, 435, 647

\bibitem[Izotov et al.(2006)]{I06} Izotov Y. I., Stasi\'nska G., Meynet G.,
Guseva N. G., Thuan T. X., 2006, \aap, 448, 955





\bibitem[Izotov et al.(2015)]{I15} Izotov Y. I., Guseva N. G., 
Fricke K. J.,  Henkel C., 2015, \mnras, 451, 2251

\bibitem[Izotov et al.(2016a)]{I16} Izotov Y. I., Orlitov\'a I., Schaerer D.,
Thuan T. X., Verhamme A., Guseva N. G.,  Worseck G., 2016a, \nat, 529, 178

\bibitem[Izotov et al.(2016b)]{I16b} Izotov Y. I., Schaerer D., Thuan, T. X., 
Worseck G., Guseva N. G., Orlitov\'a I., Verhamme A., 2016b, \mnras, 461, 3683

\bibitem[Izotov et al.(2016c)]{I16c} Izotov Y. I., Guseva N. G., 
Fricke K. J.,  Henkel C., 2016c, \mnras, 462, 4427

\bibitem[Izotov et al.(2017)Izotov, Thuan \& Guseva]{I17} Izotov Y. I., 
Thuan T. X., Guseva N. G., 2017, \mnras, 471, 548

\bibitem[Izotov et al.(2018)]{I18} Izotov Y. I., Schaerer D.,
Worseck G., Guseva N. G., Thuan, T. X., Verhamme A., Orlitov\'a I., Fricke K. J,
 2018, \mnras, 474, 4514

\bibitem[Jaskot \& Oey(2013)]{JO13} Jaskot A. E., Oey M. S.,
2013, \apj, 766, 91

\bibitem[Jaskot \& Oey(2014)]{JO14} Jaskot A. E., Oey M. S.,
2014, \apj, 791, L19

\bibitem[Kauffmann et al.(2003)]{K03} Kauffmann G. et al.,
2003, \mnras, 341, 33

\bibitem[Kennicutt(1998)]{K98} Kennicutt R. C., Jr.,
1998, \araa, 36, 189


\bibitem[Khaire et al.(2016)]{K16} Khaire V., Srianand R., Choudhury T. R.,
Gaikwad P., 2016, \mnras, 457, 4051



\bibitem[Laigle et al.(2016)]{La16} Laigle C. et al, 2016, \apjs, 224, 24


\bibitem[Leitet et al.(2013)]{L13} Leitet E., Bergvall N., Hayes M., 
Linn\'e S., Zackrisson E., 2013, \aap, 553, A106





\bibitem[Leitherer et al.(2016)]{L16} Leitherer C., Hernandez S., 
Lee J. C., Oey M. S., 2016, \apj, 823, L64



\bibitem[Madau \& Haardt(2015)]{Madau15} Madau P., Haardt F., 
2015, \apjl, 813, L8



\bibitem[Mitra et al.(2013)Mitra, Ferrara \& Choudhury]{M13} 
Mitra S., Ferrara A., Choudhury T. R., 2013, \mnras, 428, L1

\bibitem[Mitra et al.(2018)Mitra, Choudhury \& Ferrara]{M18} 
Mitra S.,  Choudhury T. R., Ferrara A., 2018, \mnras, 473, 1416




\bibitem[Naidu et al.(2018)]{N18} Naidu R. P., Forrest B., Oesch P. A., 
Tran K.-V. H., Holden B. P., 2018, \mnras, in press; preprint arXiv:1804.06845

\bibitem[Nakajima \& Ouchi(2014)]{NO14} Nakajima K.,  Ouchi M.,
2014, \mnras, 442, 900







\bibitem[Ouchi et al.(2009)]{O09} Ouchi M. et al., 2009, \apj, 706, 1136




\bibitem[Parsa et al.(2018)Parsa, Dunlop \& McLure]{P18} Parsa S., Dunlop J.S.,
McLure R. J., 2018, \mnras, 474, 2904


\bibitem[Paulino-Afonso et al.(2018)]{PA18} Paulino-Afonso A. et al., 2018,
\mnras, 476, 5479





\bibitem[Rivera-Thorsen et al.(2017)]{RT17} Rivera-Thorsen T. E. et al., 
2017, \aap, 608, L4

\bibitem[Robertson et al.(2013)]{R13} Robertson B. E. et al.,
2013, \apj, 768, 71

\bibitem[Robertson et al.(2015)]{Robertson15} Robertson B.~E., 
Ellis R.~S., Furlanetto S.~R., Dunlop J.~S., 2015, \apjl, 802, L19 

\bibitem[Rodrigo et al.(2013)Rodrigo, Solano \& Bayo]{Rod13}
Rodrigo C., Solano E., Bayo A., 2013, The SVO Filter Profile Service, 
http://ivoa.net/documents/Notes/SVOFPS/index.html


\bibitem[Rutkowski et al.(2017)]{R17} Rutkowski M. G. et al., 2017, \apj, 
841, L27



\bibitem[Schaerer et al.(2016)]{S16} Schaerer D., Izotov Y. I., Verhamme A.,
Orlitov\'a I., Thuan T. X., Worseck G., Guseva, N. G., 2016, \aap, 591, L8


\bibitem[Shapley et al.(2016)]{Sh16} Shapley A. E., Steidel C. C., 
Strom A. L., Bogosavljevi\'c M., Reddy N. A., Siana B. Mostardi R. E., 
Rudie G. C., 2016, \apj, 826, L24




\bibitem[Stasi\'nska et al.(2015)]{S15} Stasi\'nska G., Izotov Y., 
Morisset C.,  Guseva N., 2015, \aap, 576, A83





\bibitem[Thuan \& Izotov(1997)]{TI97} Thuan T. X., Izotov Y. I.,
1997, \apj, 477, 661


\bibitem[Trebitsch et al.(2017)]{T17} Trebitsch M., Blaizot J., Rosdahl J.,
Devriendt J., Slyz A., 2017, \mnras, 470, 224


\bibitem[Vanzella et al.(2010)]{V10} Vanzella E. et al., 
2010, \apj, 725, 1011 


\bibitem[Vanzella et al.(2012)]{V12} Vanzella E. et al., 2012, \apj, 751, 70 

\bibitem[Vanzella et al.(2015)]{Va15} Vanzella E. et al., 2015, \aap, 576, A116 

\bibitem[Vanzella et al.(2018)]{Va18} Vanzella E. et al., 2018, \mnras, 476, 
L15 

\bibitem[Verhamme et al.(2015)]{V15}  Verhamme A., Orlitov\'a I., 
Schaerer D., Hayes M., 2015, \aap, 578, A7

\bibitem[Verhamme et al.(2017)]{V17} Verhamme A., Orlitov\'a I., Schaerer D., 
Izotov Y., Worseck G., Thuan T. X., Guseva N., 2017, \aap, 597, A13



\bibitem[Wise \& Chen(2009)]{WC09} Wise J. H., Cen R.,
2009, \apj, 693, 984

\bibitem[Wise et al.(2014)]{W14} Wise J.~H., Demchenko 
V.~G., Halicek M.~T., Norman M. L., Turk M. J., Abel T., Smith B. D., 
2014, \mnras, 442, 2560 


\bibitem[Worseck et al.(2016)]{W16} Worseck G., Prochaska J. X., Hennawi J. F.,
McQuinn M., 2016, \apj, 825, 144

\bibitem[Wright(2006)]{W06} Wright E. L., 2006, \pasp, 118, 1711

\bibitem[Yajima et al.(2011)Yajima, Choi \& Nagamine]{Y11} 
Yajima H., Choi J.-H., Nagamine K., 2011, \mnras, 412, 411


\bibitem[Yang et al.(2017)]{Y17} Yang H. et al., 2017, \apj, 844, 171


\end{thebibliography}



\bsp	
\label{lastpage}
\end{document}